\documentclass[aps,prb,showpacs,showkeys]{revtex4}
\begin{document}
\title{ Theory of Magnetic Polaron }

%%%%%%%%%%%%%%%%%%%%%%%%%%%%%%%%%%%%%%%%%%%%%%%%%%%%%%%%%%%%%%%%%
%
%%%%%%%%%%%%%%%%%%%%%%%%%%%%%%%%%%%%%%%%%%%%%%%%%%%%%%%%%%%%%%%%%
%
\author{A.\ L.\ Kuzemsky }
\affiliation{Bogoliubov Laboratory of Theoretical Physics,\\
 Joint Institute for Nuclear Research, 141980 Dubna, Moscow Region, Russia}
\email{kuzemsky@thsun1.jinr.ru}
\date{\today}
%\date{}
%
%\maketitle
%
%
%%%%%%%%%%%%%%%%
\begin{abstract}
%%%%%%%%%%%%%%%%
%
The concept of magnetic polaron is analysed and developed to
elucidate the nature of itinerant charge carrier states in
magnetic semiconductors and similar complex magnetic materials.
By contrasting the scattering and bound states of carriers within
the $s-d$ exchange model, the nature of bound states at finite
temperatures is clarified. The free magnetic polaron at certain
conditions is realized as a bound state of the carrier (electron
or hole) with the spin wave. Quite generally, a self-consistent
theory of a magnetic polaron is formulated within a
nonperturbative many-body approach, the Irreducible Green
Functions (IGF) method  which is used to describe the
quasiparticle many-body dynamics at finite temperatures. Within
the above many-body approach we elaborate a self-consistent
picture of   dynamic behavior of two interacting subsystems, the
localized spins and  the itinerant charge carriers. In particular,
we show that the relevant generalized mean fields emerges
naturally within our formalism. At the same time, the correct
separation of elastic scattering corrections permits one to
consider the damping effects (inelastic scattering corrections)
in the unified and coherent fashion. The damping of magnetic
polaron state, which is quite different from the damping of the
scattering states, finds a natural interpretation within the
present self-consistent scheme.
%
%%%%%%%%%%%%%%
\end{abstract}
%%%%%%%%%%%%%%
%
\maketitle
%%%%%%%%%%%%%%%%%%%%%%
%
%
%
%%%%%%%%%%%%%%%%%%%%%%
\section{Introduction}
%%%%%%%%%%%%%%%%%%%%%%
%
%
%
%
The properties of itinerant charge carriers in complex magnetic
materials are at the present time of much interest. The magnetic
polaron problem is of particular interest because  one can study
how a magnetic ion  subsystem influences electronic properties of
complex magnetic materials. Recently, semiconducting ferro- and
antiferromagnetic compounds have been studied very
extensively~\cite{mauger86,wolf88,dietl94,dietl02,konig03}.
Substances which we refer   to as magnetic semiconductors,
occupy  an intermediate position between magnetic metals and
magnetic dielectrics. Magnetic semiconductors are characterized
by the existence of two well defined subsystems, the system of
magnetic moments which are localized at lattice sites, and a band
of itinerant or conduction carriers (conduction electrons or
holes). Typical examples are the Eu-chalcogenides, where the local
moments arise  from 4f electrons of the Eu ion, and the spinell
chalcogenides containing $Cr^{3+}$ as a magnetic ion. There is
experimental evidence of a substantial mutual influence of spin
and charge subsystems in these compounds. This is possible due to
the $sp-d(f)$ exchange interaction of the localized spins and
itinerant charge carriers. An itinerant carrier perturbs the
magnetic lattice and is perturbed by the spin waves. It was shown
that the effects of the $sp-d$ or $s-f$
exchange~\cite{zener53,rk54,kasuya56,kasuya59,kasuya66}, as well
as the $sp-d(f)$ hybridization~\cite{larson88}, the
electron-phonon interaction and disorder effects contributed to
essential physics of these compounds and various anomalous
properties are found. In these phenomena, the itinerant charge
carriers play an important role and many of these anomalous
properties may be attributed to the $sp-d(f)$ exchange
interaction~\cite{kasuya66,yoshi91}. As a result, an electron
travelling through a ferromagnetic crystal will in general couple
to the magnetic subsystem. From the quantum mechanics point of
view this means that the wave function of the electron would
depend not only upon the electron coordinate but upon the state
of the spin system as well. Recently, further attempts have been
made to study and exploit carriers which are exchange-coupled to
the localized spins~\cite{ohno99,ohno00,ohno01,sato01,katay01}.
The effect of carriers on the magnetic ordering temperature is now
found to be very strong in diluted magnetic semiconductors
(DMS)~\cite{ohno99,dietl02}. Diluted magnetic semiconductors are
mixed crystals in which magnetic ions (usually $Mn^{++}$) are
incorporated in a substitutional position of the host ( typically
a II-VI or III-V ) crystal lattice. The diluted magnetic
semiconductors offer a unique possibility for a gradual change of
the magnitude and sign of exchange interaction by means of
technological control of carrier concentration and band parameters.\\
It was Kasuya~\cite{kasuya56,kasuya66,lyons66,kasu70,hirst78} who
first  clarified that the s-f interaction works differently in
magnetic semiconductors and in metals~\cite{smit66}. The effects
of the sp-d(f) exchange on the ferromagnetic state of a magnetic
semiconductor were discussed
in~\cite{wolfram62,white68,balten70,kasuya72,nag74}. It was shown
that the effects of the $sp-d(f)$ exchange interaction are of a
more variety in the magnetic semiconductors~\cite{kasuya66} than
in the magnetic metals~\cite{yoshi57,yoshid57,smit66}, because in
the former there are more parameters which can change over wide
ranges~\cite{kasuya66,lyons66,kasu70,hirst78}. The state of
itinerant charge carriers may be greatly modified  due to the
scattering on the localized
spins~\cite{rys67,sinkk80,stubb80,sinkk83}. Interaction with the
subsystem of localized spins leads to renormalization of bare
states and the scattering and bound state regimes may
occur~\cite{kasu68,kasuy70}. Along with with the scattering
states, an additional dressing effect due to the sp-d(f) exchange
interaction can exist  in some of these
materials~\cite{molnar67,emin80,stubb80}. To some extent, the
interaction of an itinerant carrier in a ferromagnet with spin
waves is analogous to the polaron problem in polar
crystals~\cite{yama58} if we can consider the electron and spin
waves to be separate subsystems~\cite{wolfram62}. Note, however,
that
the magnetic polaron differs from the ordinary polaron in a few important points~\cite{smit65,kasuya72,nag74,emin80,nag92}.\\
To describe this situation a careful analysis of the state of
itinerant carriers in complex magnetic materials~\cite{emin88} is
highly desirable. For this aim a few  model approaches have been
proposed.  A basic model is a combined spin-fermion model  ( SFM
) which includes interacting spin and charge
subsystems\cite{rys67,mattis68,kasuya72,kuzem85,kuzem99}. \\
The problem of   adequate physical description  of itinerant
carriers ( including a self-trapped state) within various types
of generalized spin-fermion models  has  intensively been studied
during the last
decades~\cite{gennes60,rys67,kuzem85,kuzem99,babcen81}. The
dynamic interaction of an itinerant electron with the spin-wave
system in a magnet has been studied by many
authors~\cite{kasuya56,wolfram62,kasuya66,kasuya72}, including
the effects of external fields~\cite{miran75}. It was shown
within the perturbation theory that the state of an itinerant
charge carrier is renormalized due to   the spin disorder
scattering. The second order perturbation treatment leads to the
lifetime of conduction electron and explains qualitatively the
anomalous temperature dependence of the electrical
resistivity~\cite{yanase68,kasuya68,kasu68,kasuy70,kasuya72}. The
polaron formation in the concentrated systems leads to giant
magneto-resistive effects in the
Eu chalcogenides~\cite{mauger86,molnar67}.\\
The concept of  magnetic polaron
in the magnetic material was discussed and analysed in Refs.~\cite{molnar67,mattis68,kas70,kasuya72,nag74}.
The future development of this concept was stimulated by many
experimental results and observations on magnetic semiconductors~\cite{molnar68,tadao70,torran72,akimot94,dietl94}.
A paramagnetic polaron in magnetic semiconductors was studied by Kasuya~\cite{kasuya72},
who argued on the basis of thermodynamics, that once electron is trapped into the spin cluster,
the spin alignment within the spin cluster increases and thus the potential to trap an electron increases.
The bound states around impurity ions of opposite charge and self-trapped carriers were discussed
by de Gennes~\cite{gennes60}. Emin~\cite{emin80} defined the self-trapped state and formulated that
\begin{quote}
"the unit comprising the self-trapped carrier and the associated
atomic deformation pattern is referred to as a polaron, with the
adjective small or large denoting whether the spatial extent of
the wave function of the self-trapped carrier is small or large
compared with the dimensions of a unit cell".
\end{quote}
In papers~\cite{yanase70,takeda70,tkasu70,kas70,kasuya76}, a set
of self-consistent equations for the self-trapped ( magnetic
polaron ) state was derived and it was shown that  the
paramagnetic polaron appeared discontinuously with decreasing
temperature. These studies were carried out for wide band
materials and the thermodynamic arguments were mainly
used~\cite{ohata77} in order to determine a stable configuration.
Some specific points of spin-polaron and exiton magnetic polaron
were discussed further in papers~\cite{kasumeh72,kasuya76,kasumeh76,umeha96,umeh98}.\\
Properties of the magnetic polaron in a paramagnetic
semiconductor were studied by Yanase~\cite{yanase72},
Kubler~\cite{kubler72} and by Auslender and
Katsnelson~\cite{auslend81}. The later authors~\cite{auslend83}
developed a detailed theory of the states of itinerant charge
carriers in ferromagnetic semiconductors in the spin-wave  (low
temperature)  region within the framework of variational approach.
The effect of the electron-phonon interaction on the self-trapped
magnetic polaron state was investigated by
Umehara~\cite{umeh81,umeh83}. A theory for self-trapped magnetic
polaron in ferromagnetic semiconductor with a narrow band was
formulated by Takeda and Kasuya~\cite{takeda75}. The electron
crystallization in antiferromagnetic semiconductors was studied
by Umehara~\cite{umeha85,umeha87}, and a \emph{dense magnetic
polaron state} was conjectured to describe
the physics of $Gd_{3-x}vS_{4}$.\\
A model of the bound magnetic polaron, i.e., electron trapped on
impurity or by vacancy~\cite{torran72} was developed in
papers~\cite{kubler75,kaski79}. The general thermodynamic  model
of the bound magnetic polaron  and its stability was considered
in Refs.~\cite{mauger83,spal86,replay86}.
Then   similar models were studied by several authors~\cite{bhatta87,ram88,dietl94}.\\
The state of a conduction electron in a ferromagnetic crystal (
magnetic polaron ) was investigated by Richmond~\cite{richmond70}
who deduced an expression for one-electron Green function.
Shastry and Mattis~\cite{mattis81} presented a detailed analysis
of the one-electron Green function at zero temperature. They
constructed an exact Green function for a single electron in a
ferromagnetic semiconductor and highlighted the crucial
differences between bound- and scattering-state contributions to
the electron spectral weight. A finite temperature
self-consistent theory of magnetic polaron  within the
Green functions approach was developed in~\cite{kuzem86}. \\
Recently,  new interest in the problem of magnetic polaron was
stimulated by the studies of magnetic and transport properties of
the low-density carrier ferromagnets, diluted magnetic
semiconductors
(DMS)~\cite{takey00,nogaku01,bhatt02,dietl02,konig03}. The
concept of the magnetic polaron, the self-trapped state of a
carrier  and spin wave,  attracts increasing attention because of
the anomalous magnetic, transport,and optical properties of
DMS~\cite{sakai00,umeha00,nogaku01,umeh02,umeha03,umeh03} and the
perovskite  manganites~\cite{emin88,calder00,cataud01}. For
example, a two-component phenomenological model, describing
polaron formation in colossal magnetoresistive compounds, has
been devised recently~\cite{salam01}. The paper~\cite{bhatt02}
includes a detailed
analysis of the polaron-polaron interaction effects in DMS.  \\
The purpose of the present work is to elucidate further the
nature of itinerant carrier states in magnetic semiconductors and
similar complex magnetic materials. An added motivation for
performing new consideration and a careful analysis of the
magnetic polaron problem arise  from the circumstance that the
various new materials were fabricated and tested, and a lot of new
experimental facts were accumulated. This paper deals with the
effects of the local exchange due to interaction of carrier spins
with the ionic spins or the $sp-d$ or $s-f$ exchange interaction
on the state of itinerant charge carriers. We develop in some
detail a many-body approach to the calculation of the
quasiparticle energy spectra of itinerant carriers so as to
understand their quasiparticle many-body dynamics. The concept of
the magnetic polaron is reconsidered and developed and the
scattering and bound states are thoroughly analysed.
In the previous papers, we set up the formalism of the method of
Irreducible Green Functions (IGF)~\cite{kuzem02}. This IGF method
allows one to describe  quasiparticle spectra with damping for
many-particle systems on a lattice with complex spectra and a
strong correlation in a very general and natural way. This scheme
differs from the traditional method of decoupling of an infinite
chain of  equations~\cite{tyab67} and permits a construction of
the relevant dynamic solutions in a self-consistent way at the
level of the Dyson equation without decoupling the chain of
equations of motion for the GFs.\\ In this paper, we apply the IGF
formalism to consider  quasiparticle spectra of charge carriers
for the lattice spin-fermion model consisting of two interacting
subsystems. The concepts of magnetic polaron and the scattering
and bound states are analysed and developed in some detail. We
consider thoroughly a self-consistent calculation of
quasiparticle energy spectra of the itinerant carriers. We are
particularly interested in how the scattering state appears
differently from the bound state
in magnetic semiconductor.\\
The key problem of most
of this work is the formation of magnetic polaron under
various conditions on the parameters of the spin-fermion model.
It is the purpose of this paper to explore more fully the  effects of
the sp-d(f) exchange interaction on the state of itinerant charge carriers in magnetic
semiconductors and similar complex magnetic materials.
%
%%%%%%%%%%%%%%%%%%%%%%%%%%%%%%%%%%%%%%%%%%%%%%%%%%%%%%%%%%%%%%%
\section{The Spin-Fermion Model}
%%%%%%%%%%%%%%%%%%%%%%%%%%%%%%%%%%%%%%%%%%%%%%%%%%%%%%%%%%%%%%%%%
%
The concept of the $sp-d$ ( or $d-f$ )
model plays an important role in the quantum theory of
magnetism\cite{kasuya59,kasuya66, lyons66,yoshi91,kuzem99}.
In this section, we consider the generalized $sp-d$ model  which
describes the localized $3d(4f)$-spins interacting with
$s(p)$-like conduction (itinerant) electrons ( or holes ) and
takes into consideration the electron-electron interaction.
 \\ The total Hamiltonian of the model ( for simplicity we shall call it the $s-d$ model)
is given by \begin{equation} \label{eq1}
 H = H_{s} + H_{s-d} + H_{d}
\end{equation}
The Hamiltonian of band electrons (or holes) is given by
\begin{equation}
\label{eq2} H_{s} = \sum_{ij} \sum_{\sigma}
t_{ij}a^{\dagger}_{i\sigma}a_{j\sigma} + \frac{1}{2}
\sum_{i\sigma} Un_{i\sigma}n_{i-\sigma}
\end{equation}
This is the Hubbard model. We adopt the notation
$$a_{i\sigma} =
N^{-1/2}\sum_{\vec k}  a_{k\sigma} \exp (i{\vec k} {\vec R_{i}})
\quad a^{\dagger}_{i\sigma} = N^{-1/2}\sum_{\vec k}
a^{\dagger}_{k\sigma} \exp (-i{\vec k} {\vec R_{i}})$$
In the case of a pure semiconductor, at low temperatures the
conduction electron band is empty and the Coulomb term $U$ is
therefore not so important. A partial occupation of the band
leads to an increase in the role of the Coulomb correlation. It
is clear that  we treat conduction electrons as s-electrons in the
Wannier representation. In doped DMS the
carrier system is the valence band p-holes.\\
 The band energy of Bloch electrons
$\epsilon_{\vec k }$ is defined as follows: $$t_{ij} =
N^{-1}\sum_{\vec k}\epsilon_{\vec k }  \exp[i{\vec k}({\vec R_{i}}
-{\vec R_{j}})],$$ where  $N$ is the number of  lattice sites.
For the tight-binding electrons in a cubic lattice we use the
standard expression for the dispersion
\begin{equation}\label{eq3}
\epsilon_{\vec k }   = 2\sum_{\alpha}t( \vec a_{\alpha})\cos(\vec
k \vec a_{\alpha}) \end{equation}  where $\vec a_{\alpha}$ denotes
the lattice vectors in a simple lattice with the inversion centre.\\
The term $H_{s-d}$ describes the interaction of the total
3d(4f)-spin  with the spin density of the itinerant
carriers\cite{larson88}
\begin{equation} \label{eq4}
H_{s-d} = -2\sum_{i}I{\vec \sigma_{i}}{\vec S_{i}} = - I
N^{-1/2}\sum_{kq}\sum_{\sigma}[S^{-\sigma}_{-q}a^{\dagger}_{k\sigma}
a_{k+q-\sigma} +
z_{\sigma}S^{z}_{-q}a^{\dagger}_{k\sigma}a_{k+q\sigma}]
\end{equation}
where sign factor $z_{\sigma}$ is given by $$z_{\sigma} = (+ or
-)\quad for \quad \sigma =  (\uparrow  or  \downarrow)$$ and
$$S^{-\sigma}_{-q} = \cases {S^{-}_{-q} &if $\sigma = +$ \cr
S^{+}_{-q} &if $\sigma = -$ \cr}$$.\\
For example, in DMS~\cite{dietl02} the local exchange coupling resulted from the $p-d$
hybridization between the Mn $d$ levels and the $p$ valence band
$I \sim V_{p-d}^2$ . In magnetic semiconductors this interaction can lead to the formation
of the bound electron-magnon ( polaron-like ) bound state due to the effective attraction of the
electron and magnon in the case of antiferromagnetic coupling ( $I  < 0$  ).\\
For the subsystem of localized spins we have
\begin{equation}
\label{eq5}
 H_{d} = -\frac{1}{2} \sum_{ij} J_{ij} \vec S_{i}\vec S_{j} =
-\frac{1}{2} \sum_{q} J_{q} \vec S_{q}\vec S_{-q}
\end{equation}
Here we use the notation
$$S_{i}^{\alpha} =
N^{-1/2}\sum_{\vec k}  S_{k}^{\alpha}  \exp (i{\vec k} {\vec
R_{i}}) \quad  S_{k}^{\alpha} = N^{-1/2}\sum_{\vec i}
 S_{i}^{\alpha} \exp (-i{\vec k} {\vec R_{i}})$$
$$ [ S_{k}^{\pm},S_{q}^{z}] = \frac{1}{N^{1/2}}\mp S_{k+q}^{\pm} \quad
[ S_{k}^{+},S_{q}^{-}] =  \frac{2}{N^{1/2}} S_{k+q}^{z}$$
$$J_{ij} =
N^{-1}\sum_{\vec k}J_{\vec k}  \exp[i{\vec k}({\vec R_{i}} -{\vec
R_{j}})],$$
 This term describes a direct exchange
interaction between the localized 3d (4f) magnetic moments at the
lattice sites i and j. In the DMS system this interaction is
rather small. The ferromagnetic interaction between the local
moments is mediated by the real itinerant carriers in the valence
band of the host semiconductor material. The carrier polarization
produces the RKKY exchange interaction of   local moments~\cite{rk54,kasuya56,yoshi57,smit66,yoshi91}
\begin{equation}
\label{eq5a}
 H_{RKKY} =  - \sum_{i\neq j} K_{ij}  \vec S_{i}\vec S_{j}
\end{equation}
We emphasize that $ K_{ij}\sim |I^2| \sim V_{p-d}^4$. To explain
this, let us remind that the microscopic model\cite{larson88}, which
contains  basic physics, is the Anderson-Kondo model~\cite{yoshi91}
\begin{eqnarray}
\label{eq5b} H  = \sum_{ij} \sum_{\sigma}
t_{ij}a^{\dagger}_{i\sigma}a_{j\sigma} - V\sum_{ij}
\sum_{\sigma}(a^{+}_{i\sigma}d_{j\sigma} + h.c.)
\nonumber \\
-E_{d}\sum_{i} \sum_{\sigma} n^{d}_{i\sigma} + \frac{1}{2}
\sum_{i\sigma} Un^{d}_{i\sigma}n^{d}_{i-\sigma}
\end{eqnarray}
For the symmetric case $U = 2E_{d}$ and for $U \gg V$
Eq.(\ref{eq5b}) can be mapped onto  the Kondo lattice model~\cite{yoshi91} ( KLM
)
\begin{equation}
\label{eq5c}
 H  = \sum_{ij} \sum_{\sigma}
t_{ij}a^{\dagger}_{i\sigma}a_{j\sigma} - \sum_{i}2I{\vec
\sigma_{i}}{\vec S_{i}}
\end{equation}
Here $I \sim \frac {4V^{2}}{E_{d}}$. The KLM may be viewed as the
low-energy sector of the initial model Eq.( \ref{eq5b} ). \\We
follow the previous treatments and take as our model Hamiltonian
 expression (\ref{eq1}). For the sake of brevity we omit in
this paper the U-term (low-concentration limit). This U-term can
be included into consideration straightforwardly (see
Ref.~\cite{kuzem99}). As stated above, the model will represent
an assembly of itinerant charge carriers in a periodic atomic
lattice. The carriers are represented by   quantized Fermi
operators. The lattice sites are occupied by the localized spins.
Thus, this model can really be  called the $\emph{spin-fermion}$
model.
%%%%%%%%%%%%%%%%%%%%%%%%%%%%%%%%%%%%%%%%%%%%%%%%%%%%%%%%%%%%%%%%%%%%%%%
\section{ Outline of the IGF Method}
%%%%%%%%%%%%%%%%%%%%%%%%%%%%%%%%%%%%%%%%%%%%%%%%%%%%%%%%%%%%%%%%%%%%%%%
In this section, we  discuss
the main ideas of the IGF approach that allows one to describe
completely  quasiparticle spectra with damping in a very natural way.\\
We  reformulated  the two-time GF method~\cite{kuzem02} to the
form which is especially adjusted  to correlated fermion systems
on a lattice and systems with complex spectra. A very important
concept of the whole method is the {\it Generalized Mean Fields}
(GMFs), as it was formulated in~\cite{kuzem02}. These GMFs have a
complicated structure for a strongly correlated case and complex
spectra, and are not reduced to the functional of  mean densities
of the electrons or spins when
one calculates excitation spectra at finite temperatures. \\
To clarify the foregoing, let us consider a retarded GF of the
form~\cite{tyab67}
\begin{equation}
\label{eq6} G^{r} = <<A(t), A^{\dagger}(t')>> = -i\theta(t -
t')<[A(t) A^{\dagger}(t')]_{\eta}>, \eta = \pm  
\end{equation}
As an introduction to the concept of IGFs, let us describe the
main ideas of this approach in a symbolic and simplified form. To
calculate the retarded GF $G(t - t')$,  let us write down the
equation of motion for it
\begin{equation}
\label{eq7} \omega G(\omega) = <[A, A^{\dagger}]_{\eta}> + <<[A,
H]_{-}\mid A^{\dagger}>>_{\omega}
\end{equation}
Here we use the notation $ <<A(t), A^{\dagger}(t')>>$ for the time-dependent GF and
$<<A \mid A^{\dagger} >>_{\omega}$ for its Fourier transform~\cite{tyab67}. The notation $[A,B]_{\eta}$
refers to commutation and anticommutation depending on the value of $\eta = \pm$.\\
The essence of the method
is as follows~\cite{kuzem02}: \\ It is based on the notion of the
{\it "IRREDUCIBLE"} parts of GFs (or the irreducible parts of the
operators, $A$ and $A^{\dagger}$, out of which the GF is
constructed) in terms of which it is possible, without recourse
to a truncation of the hierarchy of equations for the GFs, to
write down the exact Dyson equation and to obtain an exact
analytic representation for the self-energy operator. By
definition, we introduce the irreducible part {\bf (ir)} of the GF
\begin{equation}
\label{eq8} ^{(ir)}<<[A, H]_{-}\vert A^{\dagger}>> = <<[A, H]_{-}
- zA\vert A^{\dagger}>>
\end{equation}
The unknown constant z is defined by the condition (or constraint)
\begin{equation}
\label{eq9} <[[A, H]^{(ir)}_{-}, A^{\dagger}]_{\eta}> = 0
\end{equation}
which is an analogue of the orthogonality condition in the Mori
formalism~\cite{mori65,hmori65}.
Let us emphasize that due to the complete equivalence of the definition of the
irreducible parts for the GFs $(^{(ir)}<<[A, H]_{-}\vert A^{\dagger}>>)$ 
and operators  $(^{(ir)}[A, H]_{-}) \equiv ([A, H]_{-})^{(ir)}$ we will use both the notation 
freely  ( $ ^{(ir)}<< A  \vert B >>$  is the same as
$<< (A)^{(ir)}  \vert B >>$ ).  A choice  
one notation over another is determined by the brevity and clarity of notation only.
From the condition (\ref{eq9}) one can find
\begin{equation}
\label{eq10} z = \frac{<[[A, H]_{-}, A^{\dagger}]_{\eta}>}{<[A,
A^{\dagger}]_{\eta}>} =
 \frac{M_{1}}{M_{0}}
\end{equation}
Here $M_{0}$ and $M_{1}$ are the zeroth and first order moments
of the spectral density. Therefore, the irreducible GFs  are
defined so that they cannot be reduced to the lower-order ones by
any kind of decoupling. It is worth  noting that the term {\it
"irreducible"} in a group theory means a representation of a
symmetry operation that cannot be expressed in terms of lower
dimensional representations. Irreducible (or connected )
correlation functions are known in statistical mechanics. In the
diagrammatic approach, the irreducible vertices are defined as
graphs that do not contain inner parts connected by the
$G^{0}$-line. With the aid of the definition (\ref{eq8})   these
concepts are expressed in terms  of retarded and advanced GFs.
The procedure extracts all relevant (for the problem under
consideration) mean-field contributions and puts them into the
generalized mean-field GF which  is defined here as
\begin{equation}
\label{eq11} G^{0}(\omega) = \frac{<[A,
A^{\dagger}]_{\eta}>}{(\omega - z)}
\end{equation}
To calculate the IGF $\quad  ^{(ir)}<<[A, H]_{-}(t),
A^{\dagger}(t')>>$ in (\ref{eq7}), we have to write the equation
of motion for it after differentiation with respect to the second
time variable $t'$. The condition of orthogonality (\ref{eq9})
removes the inhomogeneous term from this equation and is a very
crucial point of the whole approach. If one introduces the
irreducible part for the right-hand side operator, as discussed
above for the ``left" operator, the equation of motion
(\ref{eq7}) can be exactly rewritten in the following form:
\begin{equation} \label{eq12} G = G^{0} + G^{0}PG^{0}
\end{equation}
The scattering operator $P$ is given by
\begin{equation}
\label{eq13} P = (M_{0})^{-1}(\quad ^{(ir)}<<[A,
H]_{-}\vert[A^{\dagger}, H]_{-}>>^{(ir)}) (M_{0})^{-1}
\end{equation}
The structure of  equation ( \ref{eq13}) enables us to determine
the self-energy operator $M$  by  analogy with the diagram
technique
\begin{equation} \label{eq14} P = M + MG^{0}P
\end{equation}
We use here the notation $ M $ for self-energy ( mass operator
in  quantum field theory ).
From the definition (\ref{eq14}) it follows that  the self-energy
operator $M$ is defined as a proper (in the diagrammatic language,
``connected") part of the scattering operator $M = (P)^{p}$. As a
result, we obtain the exact Dyson equation for the thermodynamic
double-time Green functions
\begin{equation} \label{eq15} G =
G^{0} + G^{0} M G
\end{equation}
The difference between $P$ and $M$ can be regarded as two
different solutions of two integral equations (\ref{eq12}) and
(\ref{eq15}). However, from  the Dyson equation (\ref{eq15}) only
the full GF  is seen to be expressed as a  formal solution of the
form
\begin{equation}
\label{eq16} G = [ (G^{0})^{-1} - M ]^{-1}
\end{equation}
Equation  (\ref{eq16}) can be regarded as an alternative form of
the Dyson equation (\ref{eq15}) and the {\it definition} of $M$
provides that the generalized mean-field GF $G^{0}$ is specified.
On the contrary , for the scattering operator $P$, instead of the
property $G^{0}G^{-1} + G^{0}M = 1$, one has the property
$$(G^{0})^{-1} - G^{-1} = P G^{0}G^{-1}$$  Thus, the { \it very functional
form} of the formal solution (\ref{eq16})   precisely determines
the difference between $P$ and $M$. \\ Thus, by introducing
irreducible parts of GF (or  irreducible parts of the operators,
out of which the GF is constructed) the equation of motion
(\ref{eq7}) for the GF can exactly be  ( but using the
orthogonality constraint (\ref{eq9})) transformed into the Dyson
equation for the double-time thermal GF (\ref{eq15}). This result
is very remarkable  because  the traditional form of the GF
method does not include this point. Notice that all quantities
thus considered are  exact. Approximations can be generated not
by truncating the set of coupled equations of motions but by a
specific approximation of the functional form of the mass
operator $M$ within a self-consistent scheme  expressing $M$ in
terms of the initial GF
$$ M \approx F[G]$$.
Different approximations are relevant to different physical
situations.\\The projection operator technique  has essentially
the same philosophy. But with using the constraint (\ref{eq9}) in
our approach we emphasize the fundamental and central role of the
Dyson equation for  calculation of single-particle properties of
many-body systems. The problem of reducing the whole hierarchy of
equations involving higher-order GFs by a coupled nonlinear set
of integro-differential equations connecting the single-particle
GF to the self-energy operator is rather nontrivial. A
characteristic feature of these equations is that  besides the
single-particle GF  they involve also higher-order GF. The
irreducible counterparts of the GFs, vertex functions, serve to
identify correctly the self-energy as
$$  M = G^{-1}_{0}  - G^{-1}$$
The integral form of the Dyson equation (\ref{eq15}) gives  $M$
the physical meaning of a nonlocal and energy-dependent effective
single-particle potential. This meaning can be verified for the
exact self-energy using the diagrammatic expansion for the
causal GF.\\
It is important to note that for the retarded and advanced GFs,
the notion of the proper part $M = (P)^{p}$ is symbolic in
nature~\cite{kuzem02}. In a certain sense, it is possible to say that
it is defined here by analogy with the irreducible many-particle
$T$-matrix. Furthermore, by analogy with the diagrammatic
technique, we can also introduce the proper part defined as a
solution to the integral equation (\ref{eq14}). These analogues
allow us to better understand  the formal structure of the Dyson
equation for the double-time thermal GF, but only in a symbolic
form . However, because of the identical form of the equations
for  GFs for all three types ( advanced, retarded, and causal ),
we can convert our calculations  to causal GF  at each stage of
calculations  and, thereby, confirm the substantiated nature of
definition (\ref{eq14})! We therefore should speak of an analogy
of the Dyson equation. Hereafter, we  drop this stipulating,
since it does not cause any misunderstanding. In a sense, the IGF
method is a variant of the Gram-Schmidt orthogonalization
procedure.\\It should be emphasized that the scheme presented
above gives just a general idea of the IGF method. A more exact
explanation why one should not introduce the approximation
already in $P$, instead of having to work out $M$, is given below
when working out the application
of the method to  specific problems.\\
The general philosophy of the IGF method is in the separation and
identification of elastic scattering effects and inelastic ones.
This latter point is quite often underestimated, and both effects
are mixed. However, as far as the right definition of
quasi-particle damping is concerned, the separation of elastic
and inelastic scattering processes is believed to be crucially
important for  many-body systems with complicated spectra and
strong interaction.   \\ From a technical point of view, the
elastic GMF renormalizations can exhibit  quite a nontrivial
structure. To obtain this structure correctly, one should
construct the full GF from the complete algebra of  relevant
operators and develop a special projection procedure for
higher-order GFs, in accordance with a given algebra. Then a
natural question arises how to select the relevant set of
operators $\{ A_{1}, A_{2}, ... A_{n} \}$   describing the
"relevant degrees of freedom". The above consideration suggests
an intuitive and heuristic way to the suitable procedure as
arising from an infinite chain of equations of motion
(\ref{eq7}). Let us consider the column
$$ \pmatrix{ A_{1}\cr  A_{2}\cr \vdots \cr  A_{n}\cr}$$
where
$$ A_{1} = A,\quad A_{2} = [A,H],\quad A_{3} = [[A,H],H], \ldots
A_{n} = [[... [A, \underbrace{H]...H}_{n}]$$
Then the most
general possible Green function can be expressed as a matrix
$$ \hat G = <<\pmatrix{
A_{1}\cr  A_{2}\cr \vdots \cr A_{n}\cr} \vert \pmatrix{
A^{\dagger}_{1}& A^{\dagger} _{2}& \ldots &  A^{\dagger}
_{n}\cr}>>$$
This generalized Green function describes the one-, two-, and
$n$-particle dynamics. The equation of motion for it includes, as
a particular case, the Dyson equation for single-particle Green
function, and the Bethe-Salpeter equation which is the equation
of motion for the two-particle Green function and which is an
analogue of the Dyson equation, etc . The corresponding reduced
equations should be extracted from the equation of motion for the
generalized GF with the aid of special techniques such as the
projection method and similar techniques. This must be a final
goal towards a real understanding of the true many-body dynamics.
At this point, it is worthwhile to underline that the above
discussion is a heuristic scheme only,  but not a straightforward
recipe. The specific method of introducing  the IGFs depends on
the form of operators $A_{n}$, the type of the Hamiltonian, and
conditions of
the problem.\\
Here a sketchy form of the IGF method is presented. The aim is to
introduce the general scheme and to lay the groundwork for
generalizations.  We  demonstrated  in~\cite{kuzem02} that the
IGF method is a powerful tool for describing the quasiparticle
excitation spectra, allowing a deeper understanding of elastic and
inelastic quasiparticle scattering effects and the corresponding
aspects of damping and finite lifetimes. In the present context,
it provides a clear link between the equation-of-motion approach
and the diagrammatic methods due to derivation of the Dyson
equation.  Moreover, due to the fact that it allows the
approximate treatment of the self-energy effects on a final stage,
it yields a systematic way of the construction of approximate
solutions.
%
%%%%%%%%%%%%%%%%%%%%%%%%%%%%%%%%%%%%%%%%%%%%%
\section{Charge and Spin Degrees of Freedom}
%%%%%%%%%%%%%%%%%%%%%%%%%%%%%%%%%%%%%%%%%%%%%
Our attention will be focused on the quasiparticle many-body dynamics of the s-d model. To
describe self-consistently the charge dynamics of the $s-d$ model,
one should take into account the full algebra of relevant
operators of the suitable "charge modes"  which are appropriate
when the goal is to describe self-consistently the quasi-particle
spectra of two interacting subsystems.\\
The simplest case is to consider a situation when a single
electron is injected into an otherwise perfectly pure and insulating magnetic
semiconductor. The behavior of charge carriers can be divided into two distinct
limits based on interrelation between the band width W and the exchange interaction I:
$$|2IS| \gg W; \quad |2IS| \ll W$$
Exact solution for the s-d model is known only in the strong-coupling limit, where the band
width is small compared to the exchange interaction. This case can be considered as a starting
point for the description of narrow band materials. The case of intermediate coupling,
when $|2IS| \simeq W$, makes  serious difficulties.\\
To understand how the itinerant charge carriers behave in a wide
range of  values of model parameters, consider the equations of
motion for the charge and spin variables.
\begin{equation}
\label{eq20}  [ a_{k \sigma},
H_{s}]_{-} =   \epsilon_{k}   a_{k \sigma}
\end{equation}
\begin{equation}\label{eq21}
[ a_{k \sigma}, H_{s-d} ]_{-} = - I N^{-1/2} \sum_{q} ( S^{-\sigma}_{-q} a_{q+k-\sigma} +
z_{\sigma}S^{z}_{-q} a_{q+k\sigma})
\end{equation}
\begin{equation}
\label{eq22}  [ S^{+}_{k}, H_{s-d}]_{-} = - I N^{-1}\sum_{pq} [
2S^{z}_{k-q}a^{\dagger}_{p\uparrow}a_{p+q\downarrow} - S^{+
}_{k-q} ( a^{\dagger}_{p\uparrow} a_{p+q\uparrow} -
a^{\dagger}_{p\downarrow} a_{p+q \downarrow})]
\end{equation}
\begin{equation}
\label{eq23}  [ S^{-}_{-k}, H_{s-d}]_{-} = - I N^{-1}\sum_{pq} [
2S^{z}_{k-q}a^{\dagger}_{p\downarrow}a_{p+q\uparrow} - S^{-}_{k-q}
( a^{\dagger}_{p\uparrow} a_{p+q\uparrow} -
a^{\dagger}_{p\downarrow} a_{p+q \downarrow})]
\end{equation}
\begin{equation}
\label{eq24}  [ S^{z}_{k}, H_{s-d}]_{-} = - I N^{-1}\sum_{pq} (
S^{+}_{k-q}a^{\dagger}_{p\downarrow}a_{p+q\uparrow} - S^{-}_{k-q}
a^{\dagger}_{p\uparrow} a_{p+q\downarrow}  )
\end{equation}
\begin{equation}
\label{eq25}  [ S^{+}_{k}, H_{d}]_{-} =   N^{-1/2}\sum_{q} J_{q}(
S^{z}_{q}S^{+ }_{k-q}  - S^{z}_{k-q}S^{+}_{q})
\end{equation}
\begin{equation}
\label{eq26}  [ S^{-}_{-k}, H_{d}]_{-} =   N^{-1/2}\sum_{q} J_{q}(
S^{z}_{-(k+q)}S^{-}_{q}  - S^{z}_{q}S^{-}_{-(k+q)})
\end{equation}
From   Eq.(\ref{eq20}) - Eq.(\ref{eq26}) it follows that the
localized  spin  and  and itinerant charge  variables are coupled.\\
We have the following kinds of  charge
$$ a_{k \sigma}, \quad a^{\dagger}_{k \sigma}, \quad n_{k \sigma} = a^{\dagger}_{k \sigma}a_{k \sigma}$$
and spin operators
$$ S^{+}_{k}, \quad S^{-}_{-k} = ( S^{+}_{k} )^{\dagger},$$
$$\sigma^{+}_{k} = \sum_{q}
a^{\dagger}_{k\uparrow}a_{k+q\downarrow} ;\quad \sigma^{-}_{k} =
\sum_{q} a^{\dagger}_{k\downarrow}a_{k+q\uparrow} $$
There are   additional combined operators
$$ b_{k\sigma} =  \sum_{q} ( S^{-\sigma}_{-q} a_{q+k-\sigma} +
z_{\sigma}S^{z}_{-q} a_{q+k\sigma}) $$
In the lattice ( Wannier ) representation the operator $b_{k\sigma}$ reads
\begin{equation}\label{eq27}
b_{i\sigma} =  ( S^{-\sigma}_{i} a_{i-\sigma} +
z_{\sigma}S^{z}_{i} a_{i\sigma})
\end{equation}
It was clearly shown in Refs.~\cite{kasuya56,kasuya59,wolfram62,rys67,yanase68} that the
calculation of the energy of itinerant carriers involves the dynamics of the ion spin system.
In the approximation of rigid ion spins~\cite{rys67}, i.e.:
$$ S^{x}_{j} = S^{y}_{j} = 0 \quad S^{z}_{j} = S $$
the energy shift of electron was estimated as
$$\Delta \varepsilon (k\sigma)\sim - I \frac{\sigma}{2} S  + \frac{S^{2}}{4}
\sum_{q\neq 0} \frac{|I_{q}|^{2}}{\epsilon (k) - \epsilon (k-q)}$$
The dynamic term was estimated as
$$\Delta \varepsilon (k\ \uparrow)\sim
\frac{S^{2}}{4}\sum_{Q\neq 0}  \frac{|I_{Q}|^{2}}{\epsilon (k) - \epsilon (k-Q)}  + \frac{S}{2N}
\frac{1}{(2\pi)^{3}} \int d^{3}q \frac{|I_{q}|^{2}N(\omega(q))}{\epsilon (k) -
\epsilon (k-q) - I<S^{z}> + Dq^{2}}$$
For the case of rare-earth metals, the electron-magnon interaction  in the $s-d$ model within a second-order
perturbation theory was studied by Liu and Davis~\cite{liu67} and by Kim~\cite{kim66,kim68} within
the Bogoliubov-Tyablikov~\cite{tyab67} GFs method.\\
To describe self-consistently the charge carrier dynamics of the
$s-d$ model within a sophisticated many-body approach, one should
take into account the full algebra of relevant operators of the
suitable "modes" (degrees of freedom) which are appropriate when
the goal is to describe self-consistently quasiparticle spectra
of two interacting subsystems. An important question in this
context is the self-consistent picture of the quasiparticle
many-body dynamics which takes into account the complex structure
of the spectra due to the interaction of the "modes". Since our
goal is to calculate the quasiparticle spectra of the itinerant
charge carriers, including bound carrier-spin states, a suitable
algebra of the relevant operators should be constructed.  In
principle, the complete algebra of the relevant "modes" should
include the spin variables too. The most full relevant set of the
operators is
$$\{a_{i\sigma},\quad S^{z}_{i},\quad S^{-\sigma}_{i},\quad S^{z}_{i} a_{i\sigma},\quad S^{-\sigma}_{i} a_{i-\sigma}\}  $$
That means that the corresponding relevant GF for interacting charge and spin degrees of freedom
should have the form
\begin{equation}\label{eq27a}
{\small
\pmatrix{ <<a_{i \sigma}  \vert a^{\dagger}_{j \sigma'}>> & << a_{i \sigma} \vert
S^{z}_{j}  >> & <<a_{i \sigma}  \vert S^{\sigma'}_{j}>> &
<<a_{i \sigma}  \vert a^{\dagger}_{j \sigma'} S^{z}_{j}>> & <<a_{i \sigma}  \vert a^{\dagger}_{j -\sigma'} S^{\sigma'}_{j}>>\cr
<<S^{z}_{i} \vert a^{\dagger}_{j \sigma'}>> & <<S^{z}_{i} \vert S^{z}_{j} >> & <<S^{z}_{i} \vert S^{\sigma'}_{j} >> &
<<S^{z}_{i} \vert a^{\dagger}_{j \sigma'}S^{z}_{j} >> & <<S^{z}_{i} \vert a^{\dagger}_{j -\sigma'}S^{\sigma'}_{j} >>\cr
<<S^{-\sigma}_{i} \vert a^{\dagger}_{j \sigma'}>> & <<S^{-\sigma}_{i} \vert S^{z}_{j} >> & << S^{-\sigma}_{i}\vert S^{\sigma'}_{j} >>&
<< S^{-\sigma}_{i} \vert a^{\dagger}_{j \sigma'}S^{z}_{j} >> & << S^{-\sigma}_{i} \vert a^{\dagger}_{j -\sigma'}S^{\sigma'}_{j} >>\cr
<<S^{z}_{i} a_{i\sigma}  \vert a^{\dagger}_{j\sigma'}>> & << S^{z}_{i}a_{i\sigma} \vert
S^{z}_{j}  >> & << S^{z}_{i}a_{i \sigma}  \vert S^{\sigma'}_{j}>> &
<<S^{z}_{i}a_{i \sigma}  \vert a^{\dagger}_{j \sigma'} S^{z}_{j}>> & << S^{z}_{i}a_{i \sigma}
\vert a^{\dagger}_{j -\sigma'} S^{\sigma'}_{j}>>\cr
<<S^{-\sigma}_{i}a_{i -\sigma} \vert a^{\dagger}_{j \sigma'}>> &<<S^{-\sigma}_{i}a_{i -\sigma} \vert S^{z}_{j} >> & << S^{-\sigma}_{i}a_{i -\sigma}\vert S^{\sigma'}_{j} >>&
<< S^{-\sigma}_{i} a_{i -\sigma} \vert a^{\dagger}_{j \sigma'}S^{z}_{j} >> & << S^{-\sigma}_{i}a_{i -\sigma} \vert a^{\dagger}_{j -\sigma'}S^{\sigma'}_{j} >>\cr}
}
\end{equation}
However, to make the problem more easy tractable, we will
consider below the shortest algebra of the relevant operators
$(a_{k \sigma},a^{\dagger}_{k \sigma}, b_{k\sigma},
b^{\dagger}_{k\sigma})$. However, this choice
requires a separate treatment of the spin dynamics.\\
Here we reproduce very briefly the description of the spin
dynamics of the s-d model for the sake of self-contained formulation.
The spin quasiparticle dynamics of the s-d model
was considered in detail in papers~\cite{kuzem85,kuzem99,kuzem04}.
We consider the double-time thermal GF of
localized spins~\cite{tyab67}  which is defined as
\begin{eqnarray}\label{eq28} \mathcal G^{+-}(k;t - t') =
<<S^{+}_{k}(t),S^{-}_{-k}(t')>> = -i\theta(t -
t')<[S^{+}_{k}(t),S^{-}_{-k}(t')]_{-}> = \nonumber\\ 1/2\pi
\int_{-\infty}^{+\infty} d\omega \exp(-i\omega t)
 \mathcal G^{+-}(k;\omega) \end{eqnarray}
The next step is to write down the equation of motion for the
GF.  To describe self-consistently the spin dynamics of the $s-d$
model, one should take into account the full algebra of relevant
operators of the suitable "spin modes"  which are appropriate
when the goal is to describe self-consistently the quasiparticle
spectra of two interacting subsystems. We   used the following
generalized matrix GF of the form~\cite{kuzem85,kuzem99,kuzem04}:
\begin{equation}\label{eq29}
\pmatrix{ <<S^{+}_{k}\vert S^{-}_{-k}>> & <<S^{+}_{k}\vert
\sigma^{-}_{-k}>> \cr <<\sigma^{+}_{k}\vert S^{-}_{-k}>> &
<<\sigma^{+}_{k}\vert \sigma^{-}_{-k}>> \cr} = \hat \mathcal G (k;\omega)
\end{equation}
Let
us consider the equation of motion for the GF $\hat \mathcal G(k;\omega)$.
By differentiation  of the GF $<<S^{+}_{k}(t) \vert B (t')>> $
with respect to the first time, $t$, we find
\begin{eqnarray}\label{eq30}
\omega<<S^{+}_{k} \vert B >>_{\omega} =
{2N^{-1/2}<S^{z}_{0}>\brace 0} + \\ \nonumber \frac {I}{N}
\sum_{pq} << S^{+}_{k-q}(a^{\dagger}_{p\uparrow}a_{p+q\uparrow} -
a^{\dagger}_{p\downarrow}a_{p+q\downarrow}) -
2S^{z}_{k-q}a^{\dagger}_{p\uparrow}a_{p+q\downarrow}\vert
B>>_{\omega} \nonumber\\  + N^{-1/2}\sum_{q} J_{q}<<(
S^{z}_{q}S^{+ }_{k-q}  - S^{z}_{k-q}S^{+}_{q})\vert B>>_{\omega} \nonumber
\end{eqnarray}
where
$$ B = {S^{-}_{-k} \brace \sigma^{-}_{-k}}
$$\\
Let us introduce by definition irreducible $(ir)$ operators as
\begin{eqnarray}\label{eq31}
(S^{z}_{q})~^{ir} = S^{z}_{q} -<S^{z}_{0}>\delta_{q,0}; \quad
(a^{\dagger}_{p+q\sigma}a_{p\sigma})~^{ir} =
a^{\dagger}_{p+q\sigma}a_{p\sigma} -
<a^{\dagger}_{p\sigma}a_{p\sigma}>\delta_{q,0}\\
( (S^{z}_{q})~^{ir}S^{+ }_{k-q}  -
(S^{z}_{k-q})~^{ir}S^{+}_{q})~^{ir} =  ((S^{z}_{q})~^{ir}S^{+
}_{k-q}  - (S^{z}_{k-q})~^{ir}S^{+}_{q}) - (\phi_{q} -
\phi_{k-q})S^{+}_{k}
\end{eqnarray}
From the condition (\ref{eq9})
$$<[(  (S^{z}_{q})~^{ir}S^{+}_{k-q}  -
(S^{z}_{k-q})~^{ir}S^{+}_{q} - ( \phi_{q} - \phi_{k-q})S^{+}_{k}
), S^{-}_{-k} ]_{-}> = 0 $$ one can find
\begin{eqnarray}\label{eq32}
\phi_{q} = \frac{2K^{zz}_{q} + K^{-+}_{q}}{2<S^{z}_{0}>}\\
K^{zz}_{q} = <(S^{z}_{q})~^{ir}(S^{z}_{q})~^{ir}>; \quad
K^{-+}_{q} = <S^{-}_{-q}S^{+}_{q}>
\end{eqnarray}
Using the definition of the mass operator Eq.(\ref{eq14} ) the
equation of motion, Eq.(\ref{eq30} ), can be exactly transformed
to the Dyson equation, Eq.(\ref{eq15} )
\begin{equation}
\label{eq33} \hat \mathcal G = \hat \mathcal G_{0} + \hat \mathcal G_{0} \hat M\hat \mathcal G
\end{equation}
Hence, the determination of the full GF $\hat G$ has been reduced
to that of $\hat \mathcal G_{0}$ and $\hat M$.
The GF matrix $\mathcal G_{0}$  in the
generalized mean field approximation reads
\begin{equation} \label{eq34}
\hat \mathcal G_{0} = R^{-1} \pmatrix{
I^{-1}N^{1/2}\Omega_{2} & \Omega _{2} N \chi^{s}_{0}\cr \Omega
_{2} N \chi^{s}_{0} &-\Omega_{1}N\chi^{s}_{0}\cr}
\end{equation}
where
\begin{equation} \label{eq35}
R =
\Omega_{1}
+ \Omega _{2}I N^{1/2} \chi^{s}_{0}
\end{equation}
The diagonal matrix elements $\mathcal G^{11}_{0}$  read
\begin{equation} \label{eq36}
<<S^{+}_{k} \vert S^{-}_{-k}>>^{0} = \frac {2 S_{z}}{ \Omega_{1}
+ 2I^{2} S_{z} \chi^{s}_{0}(k,\omega ) }
\end{equation}
where
\begin{eqnarray} \label{eq37} \Omega_{1} = \omega -
\frac {<S^{z}_{0}>}{N^{1/2}} ( J_{0} - J_{k} ) -  N^{-1/2}
\sum_{q}( J_{q} - J_{q-k} )\frac{2K^{zz}_{q} +
K^{-+}_{q}}{2<S^{z}_{0}>} -  I ( n_{\uparrow} - n_{\downarrow})\\
\Omega_{2} =
\frac {2<S^{z}_{0}>I}{N }\\
\chi ^{s}_{0}(k,\omega) = N^{-1} \sum_{p} \frac {
(f_{p+k\downarrow} - f_{p\uparrow})}{\omega^{s}_{p,k}}
\end{eqnarray}
Here the notation was used
\begin{eqnarray}  \label{eq38}
\omega^{s}_{p,k} = (\omega + \epsilon_{p }  - \epsilon_{p+k }  - \Delta_{I} ) \\
 \Delta_{I}  = 2I S_{z}   \nonumber
\end{eqnarray}
$$n_{\sigma} = \frac{1}{N}\sum_{q}  <a^{\dagger}_{q\sigma}a_{q\sigma}>  =
\frac{1}{N} \sum_{q} f_{q\sigma}
=\sum_{q}(\exp(\beta \varepsilon(q\sigma)) + 1)^{-1} $$
$$ \varepsilon(q \sigma) = \epsilon_{q }  -z_{\sigma}I  S_{z}  $$

$$ \bar n = \sum  ( n_{\uparrow} + n_{\downarrow}); \quad 0 \le
\bar n \le 2 $$
$$ S_{z} = N^{-1/2}<S^{z}_{0}>  $$
We  assume  then that the local exchange parameter I = 0. In this limiting case we have
\begin{equation} \label{eq39}
<<S^{+}_{k} \vert S^{-}_{-k}>>^{0} = \frac {2 S_{z}}{   \omega -
S_{z} ( J_{0} - J_{k} ) -  \frac {1}{2N S_{z}} \sum_{q}( J_{q} -
J_{q-k} )(2K^{zz}_{q} + K^{-+}_{q})  }
\end{equation}
 The spectrum of quasiparticle
excitations of localized spins without damping follows from the
poles of the generalized mean-field GF (\ref{eq39})
\begin{equation} \label{eq40}
    \omega(k) =
S_{z} ( J_{0} - J_{k} ) +  \frac {1}{2N S_{z}} \sum_{q}( J_{q} -
J_{q-k} )(2K^{zz}_{q} + K^{-+}_{q})
\end{equation}
It is seen that due to the correct definition of generalized mean
fields we get the result for the localized spin Heisenberg
subsystem which includes both the simplest spin-wave result and
the result of Tyablikov decoupling as  limiting cases. In the
hydrodynamic limit $k \rightarrow 0$, $\omega \rightarrow 0$   it
leads to the dispersion law $\omega(k) = Dk^{2}$.\\ The exchange
integral $J_{k}$ can be written in the following way:
\begin{equation} \label{eq41}
J_{k} = \sum_{i} \exp { (-i \vec k \vec R_{i})} J( |\vec R_{i}|)
\end{equation}
The expansion in small $\vec k $ gives~\cite{kuzem04}
\begin{eqnarray} \label{eq42}
<<S^{+}_{k} \vert S^{-}_{-k}>>^{0} = \frac {2  S_{z} }{ \omega - \omega(k)} \\ \nonumber
    \omega(k\rightarrow 0) =
\Bigl ( S_{z} ( J_{0} - J_{k} ) +  \frac {1}{2N S_{z}} \sum_{q}(
J_{q} - J_{q-k} )(2K^{zz}_{q} + K^{-+}_{q})\Bigr )   \simeq Dk^{2}\\
\nonumber  = \Bigl (  \frac {S_{z}}{2} \eta_{0} + \frac{N}{2 
S_{z}^{2}}\sum_{q} \eta_{q}(2K^{zz}_{q} + K^{-+}_{q})  \Bigr ) k^{2}\\
\nonumber \eta_{q} =  \sum_{i} (\vec k \vec R_{i})^{2} J( |\vec
R_{i}|)\exp { (-i \vec q \vec R_{i})}
\end{eqnarray}
It is easy to analyse  the quasiparticle spectra of the $(s-d)$
model in the case of  nonzero coupling I. The full generalized
mean field GFs can be rewritten as
\begin{equation} \label{eq43}
<<S^{+}_{k} \vert S^{-}_{-k}>>^{0} = \frac {2 S_{z}}{   \omega -
Im - S_{z} ( J_{0} - J_{k} ) -  \frac {1}{2N S_{z}} \sum_{q}(
J_{q} - J_{q-k} )(2K^{zz}_{q} + K^{-+}_{q}) + 2I^{2}S_{z}
\chi^{s}_{0} (k,\omega)}
\end{equation}
\begin{equation} \label{eq44}
<<\sigma^{+}_{k} \vert \sigma^{-}_{-k}>>^{0} = \frac{\chi^{s}_{0}
(k,\omega )}{1 - I_{eff}(\omega) \chi^{s}_{0} (k,\omega)}
\end{equation}
Here the notation was used
$$ I_{eff} =\frac{2I^{2} S_{z}}{\omega - Im }; \quad m =
( n_{\uparrow} - n_{\downarrow}) $$
The precise significance of this description of spin quasiparticle dynamics appears in the next sections.
%
%
%
%%%%%%%%%%%%%%%%%%%%%%%%%%%%%%%%%%%%%%%%%%%%%%%%%%%%%%%%%%%%%%%%%%%%%%%%%%%%%%%%%%%%%%%
%%%%%%%%%%%%%%%%%%%%%%%%%%%%%%%%%%%%%%%%%%%%%%%%%%%%%%%%%%%%%%%%%%%%%%%%%%%%%%%%%%%%%%%%%
%
%
%%%%%%%%%%%%%%%%%%%%%%%%%%%%%%%%%%%%%%%%%%%%%%%%%%%%%%%%%%%
\section{Charge dynamics of the $s-d$ model. Scattering Regime}
%%%%%%%%%%%%%%%%%%%%%%%%%%%%%%%%%%%%%%%%%%%%%%%%%%%%%%%%%%%%%
In order to discuss the charge quasiparticle dynamics of the $s-d$ model, we can use
the whole development in Section 3. The concept of a magnetic polaron requires that we should
have also precise knowledge about the scattering charge states. By contrasting the bound and
scattering state regime, the properties of itinerant charge carriers and their quasiparticle many-body
dynamics can be substantially clarified.\\
We consider again the double-time thermal GF of
charge operators~\cite{tyab67}  which is defined as
\begin{eqnarray}\label{eq45} g_{k\sigma}(t - t') =
<<a_{k \sigma}(t),a^{\dagger}_{k \sigma}(t')>> = -i\theta(t -
t')<[a_{k \sigma}(t),a^{\dagger}_{k \sigma}(t')]_{+}> = \nonumber\\ 1/2\pi
\int_{-\infty}^{+\infty} d\omega \exp(-i\omega t)
g_{k\sigma}(\omega)
\end{eqnarray}
To describe the quasiparticle charge dynamics or dynamics of
carriers of the $s-d$ model  self-consistently, we should
consider the equation of motion for the  GF  $g$:
\begin{eqnarray}\label{eq46}
\omega <<a_{k \sigma}  | a^{\dagger}_{k \sigma}>>_{\omega} = 1 +
\epsilon_{k } <<a_{k \sigma}  | a^{\dagger}_{k \sigma}>>_{\omega} - \nonumber \\
I N^{-1/2} \sum_{q} <<( S^{-\sigma}_{-q} a_{q+k-\sigma} +
z_{\sigma}S^{z}_{-q} a_{q+k\sigma})|a^{\dagger}_{k \sigma}>>_{\omega} = I N^{-1/2} <<b_{k \sigma}  | a^{\dagger}_{k \sigma}>>_{\omega}
\end{eqnarray}
Let us introduce by definition irreducible $(ir)$  spin operators as
\begin{eqnarray}\label{eq47}
(S^{z}_{q})~^{ir} = S^{z}_{q} - <S^{z}_{0}>\delta_{q,0} \nonumber \\
(S^{ \sigma}_{q})~^{ir} = S^{ \sigma}_{q} - <S^{ \sigma}_{0}>\delta_{q,0} = S^{ \sigma}_{q}
\end{eqnarray}
By this definition we suppose that there is  a long-range magnetic order in the system under
consideration with the order parameter $<S^{z}_{0}>$. The irreducible operator for the transversal
spin components coincides with the initial operator.\\
Equivalently, one can write down  by definition irreducible GFs:
\begin{eqnarray}\label{eq48}
 (~^{ir}<<  S^{-\sigma}_{-q} a_{q+k-\sigma} | a^{\dagger}_{k \sigma}>>_{\omega}) =
<<  S^{-\sigma}_{-q} a_{q+k-\sigma} | a^{\dagger}_{k \sigma}>>_{\omega} \nonumber \\
 (~^{ir}<<S^{z}_{-q} a_{q+k\sigma} |a^{\dagger}_{k \sigma}>>_{\omega}) =
 <<S^{z}_{-q} a_{q+k\sigma} |a^{\dagger}_{k \sigma}>>_{\omega} -
 <S^{z}_{0}>\delta_{q,0} <<a_{k \sigma}  | a^{\dagger}_{k \sigma}>>_{\omega}
\end{eqnarray}
Then the equation of motion for the GF $g_{k\sigma}(\omega) $ can
be exactly transformed to the following form:
\begin{equation}\label{eq49}
( \omega - \varepsilon(k \sigma) )<<a_{k \sigma}  | a^{\dagger}_{k \sigma}>>_{\omega} +
I N^{-1/2} <<C_{k\sigma} \vert a^{\dagger}_{k \sigma}>> = 1
\end{equation}
Here the notation was used
\begin{equation}\label{eq50}
C_{k\sigma} = b~^{ir}_{k\sigma}  =  \sum_{q} \left( S^{-\sigma}_{-q} a_{q+k-\sigma} +
z_{\sigma}(S^{z}_{-q})~^{ir} a_{q+k\sigma}\right )
\end{equation}
Following the IGF strategy we should perform the differentiation
of the higher-order GFs on the second time $t'$ and introduce the
irreducible GFs ( operators) for the "right" side . Using this
approach the   equation of motion, Eq.(\ref{eq32}),  can be
exactly transformed into the Dyson equation Eq.(\ref{eq15})
\begin{equation}
\label{eq51}
  g_{k\sigma}(\omega) =
g_{k\sigma}^{0}(\omega) +
g_{k\sigma}^{0}(\omega) M_{k\sigma}(\omega) g_{k\sigma}(\omega)
\end{equation}
where
\begin{equation}
\label{eq52}
g_{k\sigma}^{0}(\omega) = <<a_{k \sigma}  | a^{\dagger}_{k \sigma}>>^{0} = (\omega - \varepsilon(k \sigma) )^{-1} \nonumber \\
\end{equation}
The mean-field GF  Eq.(\ref{eq52}) contains all the  mean-field
renormalizations or elastic scattering corrections. The inelastic
scattering corrections, according to  Eq.(\ref{eq51}), are
separated to the mass operator $M_{k\sigma}(\omega)$.
Here the mass operator has the following exact representation (scattering regime):
\begin{eqnarray}
\label{eq53}
M_{k\sigma}(\omega) =  M^{e-m}_{k\sigma}(\omega) \nonumber \\
 = {I^2 \over N} \sum_{qs} \Bigl (
(^{(ir)}<<S^{-\sigma}_{-q}
a_{k+q-\sigma}|S^{\sigma}_{s}a^{+}_{k+s-\sigma}>>^{(ir),p}) +
\\ (^{(ir)}<<S^{z}_{-q}
a_{k+q\sigma}|S^{z}_{s}a^{+}_{k+s\sigma}>>^{(ir),p}) \Bigl ) \nonumber
\end{eqnarray}
To calculate the mass operator $M_{k\sigma}(\omega)$, we express
the GF in terms of the correlation functions. In order to
calculate the mass operator self-consistently, we shall use
approximation of two interacting modes for $M^{e-m}$. Then the
corresponding expression  can be written as
\begin{eqnarray}
\label{eq54}
M^{e-m}_{k\sigma}(\omega) =
\frac{I^{2}}{N} \sum_{q}
\int
\frac {d\omega_{1} d\omega_{2}}
{\omega - \omega_{1} - \omega_{2}}
F_{1}(\omega_{1},\omega_{2})  \nonumber \\
\Bigl ( g_{k+q,-\sigma}(\omega_{2})({-1 \over \pi}Im
<<S^{\sigma}_{-q} \vert
S^{-\sigma}_{q}>>_{\omega_{1}}) + g_{k+q,\sigma}(\omega_{2})
({-1 \over \pi}Im
<<(S^{z}_{q})~^{ir} \vert
(S^{z}_{-q})~^{ir}>>_{\omega_{1}}) \Bigr )
\end{eqnarray}
where
$$F_{1}(\omega_{1},\omega_{2}) = (1 + N(\omega_{1}) -
f(\omega_{2}))$$
$$N(\omega(k)) = [\exp (\beta \omega(k)) - 1]^{-1}$$
Equations  (\ref{eq51}) and   (\ref{eq54})  form a closed
self-consistent system of equations for one-fermion GF of  the
carriers for the s-d model in the scattering regime. It clearly
shows that the charge quasiparticle dynamics couples intrinsically
with the spin quasiparticle dynamics in a self-consistent way.\\
To find explicit expressions for the mass operator,
Eq.(\ref{eq38}), we choose for the first iteration step in its
r.h.s. the following trial expressions:
\begin{eqnarray}
\label{eq55}
g_{k\sigma}(\omega) = \delta(\omega - \varepsilon(k\sigma)) \\
{-1 \over \pi} Im <<S^{\sigma}_{q} \vert
S^{-\sigma}_{-q}>> \approx z_{\sigma}
(2 S_{z} ) \delta(\omega - z_{\sigma} \omega(q))
\end{eqnarray}
Here $\omega(q)$ is given by the expression Eq.(\ref{eq40}).
Then we obtain
\begin{eqnarray}
\label{eq56}
M^{e-m}_{k\uparrow}(\omega) =
\frac{2I^{2}<S^{z}_{0}>}{N^{3/2}} \sum_{q}
\frac {f_{k+q,\downarrow} + N(\omega(q))}
{\omega - \varepsilon(k+q,\downarrow) - \omega(q)}~; \nonumber \\
M^{e-m}_{k\downarrow}(\omega) =
\frac{2I^{2}<S^{z}_{0}>}{N^{3/2}} \sum_{q}
\frac {1- f_{k-q,\uparrow} + N(\omega(q))}
{\omega - \varepsilon(k-q,\uparrow) - \omega(q)}
\end{eqnarray}
This result was written for the low temperature region  when one
can drop the contributions from the dynamics of longitudinal spin
GF. The last is essential at high temperatures and in some
special cases. The obtained formulas generalize the
zero-temperature calculations of Davis and Liu~\cite{liu67} and
the  approach of papers~\cite{rys67,kim66,kim68}. The numerical
calculations of the typical behaviour of the real and imaginary
parts of the self-energy in generalized Born approximation were
carried out in Refs.~\cite{sinkk80,sinkk83}.
%%%%%%%%%%%%%%%%%%%%%%%%%%%%%%%%%%%%%%%%%%%%%%%%%%%%%%%%%%%%%%%%%
%%%%%%%%%%%%%%%%%%%%%%%%%%%%%%%%%%%%%%%%%%%%%%%%%%%%%%%%%%%%%%%%%%%%%%%%%%%%%%%%%%%
\section{Charge Quasiparticle Dynamics of the $s-d$ Model. Bound State Regime.}
%%%%%%%%%%%%%%%%%%%%%%%%%%%%%%%%%%%%%%%%%%%%%%%%%%%%%%%%%%%%%%%%%%%%%%%%%%%%%%%%%%%%%
In this section, we further discuss  the spectrum of charge
carrier excitations in the $s-d$ model and describe bound state
regime. As previously, consider the double-time thermal GF of
charge operators $<<a_{k \sigma}(t),a^{\dagger}_{k
\sigma}(t')>>$. The next step is to write down the equation of
motion for the GF $g$:
\begin{equation}\label{eq57}
( \omega - \varepsilon(k \sigma) )<<a_{k \sigma}  | a^{\dagger}_{k \sigma}>>_{\omega} +
I N^{-1/2} <<C_{k\sigma} \vert a^{\dagger}_{k \sigma}>>_{\omega} = 1
\end{equation}
We also have
\begin{equation}\label{eq59}
( \omega - \varepsilon(k \sigma) )<<a_{k \sigma}  | C^{\dagger}_{k \sigma}>>_{\omega} +
I N^{-1/2} <<C_{k\sigma} \vert C^{\dagger}_{k \sigma}>>_{\omega} = 0
\end{equation}
It   follows from   Eqs.(\ref{eq57})and (\ref{eq59}) that to take
into account both the regimes, scattering and bound state,
properly, we should treat the operators $a_{k \sigma}  ,
a^{\dagger}_{k \sigma}$ and $ C_{k\sigma}, C^{\dagger}_{k\sigma}
$ on the equal footing. That means that one should consider the
new relevant operator,  a kind of 'spinor' ${a_{k\sigma}\choose
C_{k\sigma}}$ ("relevant degrees of freedom") to construct a
suitable Green function. Thus,  according to the IGF strategy, to
describe the bound state regime properly, contrary to the
scattering regime, one should consider the generalized matrix GF
of the form
\begin{equation}\label{eq60}
\pmatrix{ <<a_{k \sigma}  \vert a^{\dagger}_{k \sigma}>>_{\omega} & << a_{k \sigma} \vert
C^{\dagger}_{k\sigma}  >>_{\omega} \cr <<C_{k\sigma} \vert a^{\dagger}_{k \sigma}>>_{\omega} &
<<C_{k\sigma} \vert C^{\dagger}_{k\sigma} >>_{\omega} \cr} = \hat G(k;\omega)
\end{equation}
Equivalently, we can do the calculations in the Wannier
representation with the matrix of the form
\begin{equation}\label{eq61}
\pmatrix{ <<a_{i \sigma}  \vert a^{\dagger}_{j \sigma}>> & << a_{i \sigma} \vert
C^{\dagger}_{j\sigma}  >> \cr <<C_{i\sigma} \vert a^{\dagger}_{j \sigma}>> &
<<C_{i\sigma} \vert C^{\dagger}_{j\sigma} >> \cr} = \hat G(ij;\omega)
\end{equation}
The form of Eq.(\ref{eq61}) is  more convenient for considering the effects of disorder.\\
Let us consider now the equation of motion for the GF $\hat G(k;\omega)$.
To write down the equation of motion for the Fourier transform of
the GF   $\hat G(k;\omega)$, we need   auxiliary equations of
motion for the following GFs of the form
\begin{eqnarray}\label{eq62}
( \omega - \varepsilon(k +q-\sigma) )<< S^{-\sigma}_{-q}a_{k + q-\sigma}  | a^{\dagger}_{k \sigma}>>_{\omega} =
-I N^{-1/2} <<S^{-\sigma}_{-q} C_{k+q-\sigma} \vert a^{\dagger}_{k \sigma}>>_{\omega} -\nonumber \\
z_{\sigma} N^{-1/2} \sum_{p} J_{p}
<<( S^{-\sigma}_{-(p+q)}S^{z}_{p} -  S^{-\sigma}_{p}S^{z}_{-(p+q)})a_{q+k-\sigma}| a^{\dagger}_{k \sigma}>>_{\omega} \nonumber \\
= -I N^{-1/2}  \sum_{p}<<S^{-\sigma}_{-q} \left( S^{\sigma}_{-p} a_{p+k+q\sigma} +
z_{-\sigma}(S^{z}_{-p})~^{ir} a_{p+k+q-\sigma}\right ) \vert a^{\dagger}_{k \sigma}>>_{\omega}
-\nonumber \\
z_{\sigma} N^{-1/2} \sum_{p} J_{p}
<<( S^{-\sigma}_{-(p+q)}S^{z}_{p} -  S^{-\sigma}_{p}S^{z}_{-(p+q)})a_{q+k-\sigma}| a^{\dagger}_{k \sigma}>>_{\omega}
\end{eqnarray}
To separate the elastic and inelastic scattering corrections, it
is convenient to introduce by definition the following set of
irreducible operators:
\begin{eqnarray}\label{eq63}
( S^{ \sigma}_{-p} S^{- \sigma}_{-q})~^{ir} =  S^{ \sigma}_{-p} S^{- \sigma}_{-q} -
< S^{ \sigma}_{q} S^{- \sigma}_{-q}>\delta_{-q,p}\\
( S^{-\sigma}_{-(p+q)}S^{z}_{p} -  S^{-\sigma}_{p}S^{z}_{-(p+q)})~^{ir} =
( S^{-\sigma}_{-(p+q)}S^{z}_{p} -  S^{-\sigma}_{p}S^{z}_{-(p+q)}) -  \nonumber\\
\left ( <S^{z}_{0}> (\delta_{p,0} -  \delta_{p,-q} ) + ( \phi_{-p} - \phi_{-(p+q)})\right ) S^{ -\sigma}_{-q})
\nonumber
\end{eqnarray}
This is the standard way of introducing the "irreducible" parts
of operators or GFs~\cite{kuzem02}. However, we are interested
here in describing  the bound electron-magnon states correctly.
Thus, the definition of the relevant generalized mean field is
more tricky for this case. It is important to note that
\emph{before} introducing the irreducible parts, Eq.(\ref{eq63}),
one has to extract from the GF $ <<S^{-\sigma}_{-q}
C_{k+q-\sigma} \vert a^{\dagger}_{k \sigma}>>$ the terms
proportional to the initial GF $ <<S^{-\sigma}_{-q}
a_{k+q-\sigma} \vert a^{\dagger}_{k \sigma}>>$. That means that we
should project the higher-order GF onto the initial
one~\cite{kuzem02}. This projection should be performed using the
spin commutation relations:
$[S^{-\sigma}_{-q}, S^{z}_{-p}  ] = z_{\sigma} N^{-1/2}S^{-\sigma}_{-(q+p)} ;\quad
$$[S^{-\sigma}_{-q}, S^{\sigma}_{-p }  ] = z_{-\sigma} 2 N^{-1/2}S^{z}_{-(q+p)}$. \\
In other words, this procedure introduces effectively the
spin-operator ordering rule into the calculations. Roughly
speaking, we should construct the relevant mean field not for
spin or electron alone, but for the complex object, the
"spin-electron", or for the operator $( S^{-\sigma}_{-q}
a_{k+q-\sigma})$. This is the crucial point of the whole
treatment, which leads to the correct definition of the
generalized mean field in which the free magnetic polaron will
propagate.
We have then
\begin{eqnarray}\label{eq64}
<< S^{-\sigma}_{-q}S^{z}_{-p} a_{k +q+p-\sigma}  | a^{\dagger}_{k \sigma}>>_{\omega} =
z_{\sigma} N^{-1/2}<< S^{-\sigma}_{-(q+p)} a_{k +q+p -\sigma}  | a^{\dagger}_{k \sigma}>>_{\omega}\\ \nonumber
+ <<S^{z}_{-p} S^{-\sigma}_{-q} a_{k +q+p -\sigma}  | a^{\dagger}_{k \sigma}>>_{\omega}\\
\label{eq65}
<< S^{z}_{-p}S^{-\sigma}_{-q} a_{k +q+p-\sigma}  | a^{\dagger}_{k \sigma}>>_{\omega} =
<< (S^{z}_{-p})~^{ir}S^{-\sigma}_{-q} a_{k +q+p-\sigma}  | a^{\dagger}_{k \sigma}>>_{\omega} \\ \nonumber
+   <S^{z}_{0}> \delta_{p,0} << S^{-\sigma}_{-q} a_{k +q -\sigma}  | a^{\dagger}_{k \sigma}>>_{\omega}\\
\label{eq66}
<< S^{-\sigma}_{-q}S^{\sigma }_{-p} a_{k +q+p \sigma}  | a^{\dagger}_{k \sigma}>>_{\omega} =
<< (S^{-\sigma}_{-q}S^{\sigma }_{-p} a_{k +q+p \sigma})~^{ir}  | a^{\dagger}_{k \sigma}>>_{\omega}\\ \nonumber
+ < S^{-\sigma}_{-q}S^{\sigma }_{q}> \delta_{p,-q} <<a_{k \sigma}  \vert a^{\dagger}_{k \sigma}>>\\
<< (S^{-\sigma}_{-(p+q)}S^{z}_{p}  - S^{-\sigma}_{p}S^{z}_{-(p+q)} )a_{k +q-\sigma}  | a^{\dagger}_{k \sigma}>>_{\omega} =
\label{eq67}
<< \left ( S^{-\sigma}_{-(p+q)} (S^{z}_{p})~^{ir}  -
S^{-\sigma}_{p}( S^{z}_{-(p+q)})~^{ir} \right )a_{k +q-\sigma}  | a^{\dagger}_{k \sigma}>>_{\omega}\\ \nonumber   +
 <S^{z}_{0}> (\delta_{p,0} - \delta_{p,-q}) << S^{-\sigma}_{-q} a_{k +q -\sigma}  | a^{\dagger}_{k \sigma}>>_{\omega}
\end{eqnarray}
Finally, by differentiation  of the GF $<<S^{-\sigma}_{-q}
a_{k+q-\sigma} (t), a^{\dagger}_{k \sigma}(0)>> $ with respect to
the first time, $t$,  and using the definition of the irreducible
parts, Eq.(\ref{eq64} ) -  Eq.(\ref{eq67} ), the equation of
motion, Eq.(\ref{eq62} ),   can be exactly transformed to the
following form:
\begin{eqnarray}\label{eq68}
(\omega + z_{\sigma}\omega(q) - \varepsilon(k+q -\sigma))<<S^{-\sigma}_{-q} a_{k+q-\sigma} \vert a^{\dagger}_{k \sigma}>>_{\omega} +
I N^{-1/2}< S^{-\sigma}_{-q}S^{\sigma }_{q}> <<a_{k \sigma}  \vert a^{\dagger}_{k \sigma}>>_{\omega} = \nonumber \\
I N^{-1/2} \sum_{p}<<S^{-\sigma}_{-p} a_{k+p-\sigma} \vert a^{\dagger}_{k \sigma}>>_{\omega} + <<A_{q}  \vert a^{\dagger}_{k \sigma}>>_{\omega}
\end{eqnarray}
where
\begin{eqnarray} \label{eq69} A_{q} = - I N^{-1/2} \sum_{p} \{ (S^{-\sigma}_{-q} S^{-\sigma}_{-p}a_{k+q+p \sigma})~^{ir}
+ z_{-\sigma} ( S^{z}_{-p} )~^{ir} S^{-\sigma}_{-q} a_{k+q+p -\sigma}\} - \nonumber \\
- z_{\sigma} N^{-1/2} \sum_{p} J_{p} \left ( S^{-\sigma}_{-(p+q)} (S^{z}_{p})~^{ir} -
S^{-\sigma}_{p}( S^{z}_{q+p} )~^{ir} \right )~^{ir} a_{k+q-\sigma} = \\
- I N^{-1/2} C_{k+q-\sigma} S^{-\sigma}_{-q} - z_{\sigma} N^{-1/2} \sum_{p} J_{p} \left ( S^{-\sigma}_{-(p+q)} (S^{z}_{p})~^{ir} -
S^{-\sigma}_{p}( S^{z}_{q+p} )~^{ir} \right )~^{ir} a_{k+q-\sigma} \nonumber
\end{eqnarray}
It is easy to see that
\begin{eqnarray}\label{eq70}
<<S^{-\sigma}_{-q} a_{k+q-\sigma} \vert a^{\dagger}_{k \sigma}>>_{\omega} +
I N^{-1/2} \frac{< S^{-\sigma}_{-q}S^{\sigma }_{q}>}{(\omega + z_{\sigma}\omega(q) - \varepsilon(k+q -\sigma))} <<a_{k \sigma}
\vert a^{\dagger}_{k \sigma}>>_{\omega} = \nonumber \\
I N^{-1/2}\frac{1}{(\omega + z_{\sigma}\omega(q) - \varepsilon(k+q -\sigma))}
\sum_{p}<<S^{-\sigma}_{-p} a_{k+p-\sigma} \vert a^{\dagger}_{k \sigma}>>_{\omega} +
 \frac{1}{(\omega + z_{\sigma}\omega(q) - \varepsilon(k+q -\sigma))} <<A_{q}  \vert a^{\dagger}_{k \sigma}>>_{\omega}
\end{eqnarray}
After summation with respect to $q $ we find
\begin{eqnarray}\label{eq71}
\{ I N^{-1/2}\sum_{q}  \frac{< S^{-\sigma}_{-q}S^{\sigma }_{q}>}{(\omega + z_{\sigma}\omega(q) - \varepsilon(k+q -\sigma))} \}
<<a_{k \sigma} \vert a^{\dagger}_{k \sigma}>>_{\omega}  + \nonumber \\
\{ 1 -  I N^{-1}\sum_{q} \frac{1}{(\omega + z_{\sigma}\omega(q) -
\varepsilon(k+q -\sigma))}\} \sum_{p}<<S^{-\sigma}_{-p} a_{k+p-\sigma} \vert a^{\dagger}_{k \sigma}>>_{\omega}  = \nonumber \\
\sum_{q}  \frac{1}{(\omega + z_{\sigma}\omega(q) - \varepsilon(k+q -\sigma))} <<A_{q}  \vert a^{\dagger}_{k \sigma}>>_{\omega}
\end{eqnarray}
Then Eq.(\ref{eq70} ) can be exactly rewritten in the following
form:
\begin{eqnarray}\label{eq72}
\sum_{q} <<S^{-\sigma}_{-p} a_{k+p-\sigma} \vert a^{\dagger}_{k \sigma}>>_{\omega} = -
\{ I N^{-1/2}\sum_{q}  \frac{< S^{-\sigma}_{-q}S^{\sigma }_{q}>}{(1 - I \Lambda_{k\sigma}(\omega))(\omega + z_{\sigma}\omega(q) - \varepsilon(k+q -\sigma))} \}
<<a_{k \sigma} \vert a^{\dagger}_{k \sigma}>>_{\omega}  + \nonumber \\
+ \sum_{q}  \frac{1}{(1 - I \Lambda_{k\sigma}(\omega))(\omega + z_{\sigma}\omega(q) - \varepsilon(k+q -\sigma))} <<A_{q}  \vert a^{\dagger}_{k \sigma}>>_{\omega}
\end{eqnarray}
where
\begin{equation} \label{eq73}
 \Lambda_{k\sigma}(\omega) = \frac{1}{N} \sum_{q}  \frac{1}{(\omega + z_{\sigma}\omega(q) - \varepsilon(k+q -\sigma))}
\end{equation}
To write down the equation of motion for the matrix GF $\hat G(k;\omega)$ Eq.(\ref{eq60} )  it is
necessary to return to the operators $C_{k \sigma}$. We find
\begin{eqnarray}\label{eq74}
I N^{-1/2}\sum_{q} \{\frac{< S^{-\sigma}_{-q}S^{\sigma }_{q}>}{(1 - I \Lambda_{k\sigma}(\omega))(\omega +
z_{\sigma}\omega(q) - \varepsilon(k+q -\sigma))}   +   \frac{<( S^{z}_{-q} )~^{ir}( S^{z}_{q} )~^{ir}> }{\omega - \varepsilon(k+q \sigma)}\}
<<a_{k \sigma} \vert a^{\dagger}_{k \sigma}>>_{\omega}  + \nonumber \\
<<C_{k \sigma} \vert a^{\dagger}_{k \sigma}>>_{\omega}   =
\sum_{q} \{ \frac{<<A_{q}  \vert a^{\dagger}_{k \sigma}>>_{\omega} }{(1 - I \Lambda_{k\sigma}(\omega))(\omega + z_{\sigma}\omega(q) - \varepsilon(k+q -\sigma)) }
+ \frac{<<B_{q}  \vert a^{\dagger}_{k \sigma}>>_{\omega}}{\omega - \varepsilon(k+q \sigma)} \}
\end{eqnarray}
\begin{eqnarray}\label{eq75}
I N^{-1/2}\sum_{q} \{\frac{< S^{-\sigma}_{-q}S^{\sigma }_{q}>}{(1 - I \Lambda_{k\sigma}(\omega))(\omega +
z_{\sigma}\omega(q) - \varepsilon(k+q -\sigma))}   +   \frac{<( S^{z}_{-q} )~^{ir}( S^{z}_{q} )~^{ir}> }{\omega - \varepsilon(k+q \sigma)}\}
<<a_{k \sigma} \vert C^{\dagger}_{k \sigma}>>_{\omega}  + \nonumber \\
<<C_{k \sigma} \vert C^{\dagger}_{k \sigma}>>_{\omega}   =
\sum_{q} \{\frac{< S^{-\sigma}_{-q}S^{\sigma }_{q}>}{(1 - I \Lambda_{k\sigma}(\omega))(\omega +
z_{\sigma}\omega(q) - \varepsilon(k+q -\sigma))}   +   \frac{<( S^{z}_{-q} )~^{ir}( S^{z}_{q} )~^{ir}> }{\omega - \varepsilon(k+q \sigma)}\} +
\nonumber \\
\sum_{q} \{ \frac{<<A_{q}  \vert C^{\dagger}_{k \sigma}>>_{\omega} }{(1 - I \Lambda_{k\sigma}(\omega))(\omega + z_{\sigma}\omega(q) - \varepsilon(k+q -\sigma)) }
+ \frac{<<B_{q}  \vert C^{\dagger}_{k \sigma}>>_{\omega}}{\omega - \varepsilon(k+q \sigma)} \}
\end{eqnarray}
where
\begin{equation} \label{eq76}
B_{q} = - I N^{-1/2}\sum_{p} [ ( S^{z}_{-q} S^{z }_{-p} a_{k+q+p \sigma})~^{ir} +
z_{\sigma}( S^{z}_{-q})~^{ir}S^{-\sigma}_{-p}a_{k+q+p -\sigma} ] =
 - z_{\sigma}I N^{-1/2}( S^{z}_{-q})~^{ir}C_{k + q\sigma}
\end{equation}
The irreducible operators Eq.(\ref{eq63} ), Eq.(\ref{eq64} ) - Eq.(\ref{eq67} ) have been
introduced in such a way that the the operators $A_{q}$ and $B_{q}$ satisfy the conditions
\begin{eqnarray}\label{eq77}
< [ A_{q}, a^{\dagger}_{k \sigma} ]_{+}> = < [ A_{q}, C^{\dagger}_{k \sigma}]_{+}> = 0 \nonumber \\
< [  B_{q},   a^{\dagger}_{k \sigma}]_{+} > = <[ B_{q}, C^{\dagger}_{k \sigma}]_{+}> = 0
\end{eqnarray}
The equations of motion, Eqs.(\ref{eq74} ) and  (\ref{eq75} ) can
be rewritten in the following form:
\begin{eqnarray}\label{eq78}
 I N^{-1/2} \chi^{b}_{k \sigma} (\omega) <<a_{k \sigma} \vert a^{\dagger}_{k \sigma}>>_{\omega} +
 <<C_{k \sigma} \vert a^{\dagger}_{k \sigma}>>_{\omega} = \nonumber \\
\sum_{q} \{ \frac{<<A_{q}  \vert a^{\dagger}_{k \sigma}>>_{\omega} }{(1 - I \Lambda_{k\sigma}(\omega))(\omega + z_{\sigma}\omega(q) - \varepsilon(k+q -\sigma)) }
\nonumber \\
+ \frac{(1 + I \Lambda_{k\sigma}(\omega))  <<B_{q}  \vert a^{\dagger}_{k \sigma}>>_{\omega}}{(1 - I \Lambda_{k\sigma}(\omega)) (\omega - \varepsilon(k+q \sigma))} \}
\end{eqnarray}
\begin{eqnarray}\label{eq79}
 I N^{-1/2} \chi^{b}_{k \sigma} (\omega) <<a_{k \sigma} \vert C^{\dagger}_{k \sigma}>>_{\omega} +
 <<C_{k \sigma} \vert C^{\dagger}_{k \sigma}>>_{\omega} = \nonumber \\
\sum_{q} \{\frac{< S^{-\sigma}_{-q}S^{\sigma }_{q}>}{(1 - I \Lambda_{k\sigma}(\omega))(\omega +
z_{\sigma}\omega(q) - \varepsilon(k+q -\sigma))}   +   \frac{(1 + I \Lambda_{k\sigma}(\omega))  <( S^{z}_{-q} )~^{ir}( S^{z}_{q} )~^{ir}> }{ (1 -
 I \Lambda_{k\sigma}(\omega))(\omega - \varepsilon(k+q \sigma))}\} \\
\sum_{q} \{ \frac{<<A_{q}  \vert C^{\dagger}_{k \sigma}>>_{\omega} }{(1 - I \Lambda_{k\sigma}(\omega))(\omega + z_{\sigma}\omega(q) - \varepsilon(k+q -\sigma)) }
+ \frac{(1 + I \Lambda_{k\sigma}(\omega)) <<B_{q}  \vert C^{\dagger}_{k \sigma}>>_{\omega}}{(1 -
I \Lambda_{k\sigma}(\omega)) (\omega - \varepsilon(k+q \sigma))} \} \nonumber
\end{eqnarray}
where
\begin{eqnarray}\label{eq80}
\chi^{b}_{k \sigma} (\omega) =
\sum_{q} \{ \frac{ < S^{-\sigma}_{-q}S^{\sigma }_{q}>}{(1 - I \Lambda_{k\sigma}(\omega))(\omega + z_{\sigma}\omega(q) - \varepsilon(k+q -\sigma)) }
\nonumber \\
+ \frac{(1 + I \Lambda_{k\sigma}(\omega)) <(S^{z}_{-q})~^{ir}(S^{z}_{q})~^{ir}> }{(1 - I \Lambda_{k\sigma}(\omega)) (\omega - \varepsilon(k+q \sigma))} \}
\end{eqnarray}
Here $\chi^{b}_{k \sigma} (\omega)$ plays the role of the generalized "susceptibility" of the
spin-electron bound states instead of the electron susceptibility $\chi ^{s}_{0}(k,\omega)$
in the scattering-state regime Eq.(\ref{eq37} )  ( see also Refs.~\cite{kuzem85,kuzem99}).
Analogously, one can write the equation for the GF $<<C_{k \sigma} \vert C^{\dagger}_{k \sigma}>> $.\\
Now we are ready to write down the equation of motion for the
matrix GF $\hat G(k;\omega) $,  Eq.(\ref{eq60} ), after
differentiation with respect to the first time, $t$.  Using the
equations of motion  (\ref{eq75}), (\ref{eq78}),  (\ref{eq79} )
and  (\ref{eq57} ), we find
\begin{eqnarray}
\label{eq81}
\hat \Omega \hat G(k;\omega) = \hat I + \sum_{p}\hat
\Phi(p) \hat D(p;\omega)
\end{eqnarray}
where
\begin{eqnarray}
\label{eq82} \quad \hat \Omega = \pmatrix{ \omega - \varepsilon(k\sigma) & IN^{1/2}
\cr IN^{1/2}\chi^{b}_{k \sigma} (\omega) &1 \cr}
\quad  \hat I = \pmatrix{ 1 & 0\cr
0& \chi^{b}_{k \sigma} (\omega)\cr}  \\
\hat D(p;\omega) =  \pmatrix{ <<A_{p}
\vert  a^{\dagger}_{k \sigma}>>_{\omega} & <<A_{p} \vert  C^{\dagger}_{k \sigma}>>_{\omega} \cr <<B_{p}
\vert  a^{\dagger}_{k \sigma}>>_{\omega} & <<B_{p} \vert  C^{\dagger}_{k \sigma}>>_{\omega} \cr}
\quad
\hat \Phi(p) = \pmatrix{ 0 & 0\cr \frac{1}{\omega^{b}_{k,p }} & \frac{1}{\Omega_{k,p }} \cr}
\end{eqnarray}
with the notation
\begin{eqnarray}
\label{eq83}
\omega^{b}_{k,q } = (1 - I \Lambda_{k\sigma}(\omega))(\omega + z_{\sigma}\omega(q) - \varepsilon(k+q -\sigma)) \\
\Omega_{k,q } = \frac{(1 - I \Lambda_{k\sigma}(\omega)) }{(1 + I \Lambda_{k\sigma}(\omega))}(\omega - \varepsilon(k+q \sigma))
\end{eqnarray}
To calculate the higher order GFs  $\hat D(p;\omega)$ in Eq.( \ref{eq81}), we
differentiate its r.h.s.   with respect to the second-time
variable (t').
After introducing the irreducible parts as discussed above , but this time for the "right" operators,
and combining both (the first- and second-time differentiated)
equations of motion, we get the "exact"( no approximation has
been made till now) "scattering" equation
\begin{equation}
\label{eq84}
 \hat G(k;\omega) = \hat G^{0}(k;\omega) +   \hat G^{0}(k;\omega)  \hat
P  \hat G^{0}(k;\omega)
\end{equation}
Here the generalized mean-field GF was defined as
\begin{equation}
\label{eq85}
 \hat G^{0}(k;\omega)  = \Omega_{k,p }^{-1}    \hat I
\end{equation}
Note  that it is possible to arrive at equation  ( \ref{eq84})
using the symmetry properties too. We have
\begin{eqnarray}\label{eq86}
 \hat G^{\dagger} \hat \Omega^{\dagger} = \hat I^{\dagger} +
 \sum_{q} \hat D^{\dagger} \hat \Phi^{\dagger}(q) \nonumber \\
 \hat G = \hat I^{\dagger} (\hat \Omega^{\dagger} )^{-1} +
\sum_{q} \hat D^{\dagger} \hat \Phi^{\dagger}(q)(\hat \Omega^{\dagger} )^{-1} \nonumber \\
\Omega \hat D^{\dagger} = \sum_{p}\hat \Phi(p) \hat P (pq ) \quad \hat G^{\dagger} =
\hat G = (\hat \Omega^{-1} \hat I ) \nonumber \\
\hat G = (\hat \Omega^{-1} \hat I )^{\dagger} +
(\hat \Omega^{-1} \hat I ) \sum_{pq} \hat I^{-1}
\hat \Phi(p)\hat P (pq )\hat \Phi^{\dagger}(q)(\hat I^{-1} )^{\dagger} (\hat \Omega^{-1} \hat I )^{\dagger}
\nonumber
\end{eqnarray}
\begin{equation}
\label{eq87}
\hat P = \hat I^{-1}\{ \sum_{pq}  \hat \Phi(p)  \hat P (pq )  \hat \Phi^{\dagger}(q)\}\hat I^{-1}
\end{equation}
\begin{equation}
\label{eq88}
\hat P (pq )  =  \pmatrix{ <<A_{p}
\vert  A^{\dagger}_{q}>> & <<A_{p} \vert  B^{\dagger}_{q}>> \cr <<B_{p}
\vert  A^{\dagger}_{q}>> & <<B_{p} \vert  B^{\dagger}_{q}>> \cr}
\end{equation}
We shall now consider the magnetic polaron state in the generalized mean field
approximation and estimate the binding energy of the magnetic polaron.
\section{ Magnetic Polaron in Generalized Mean Field}
From the definition, Eq.( \ref{eq85}), the generalized mean-field
GF matrix reads
\begin{equation} \label{eq89}
\hat G^{0}(k;\omega) = \pmatrix{ <<a_{k \sigma}  \vert a^{\dagger}_{k \sigma}>>^{0} & << a_{k \sigma} \vert
C^{\dagger}_{k\sigma}  >>^{0} \cr <<C_{k\sigma} \vert a^{\dagger}_{k \sigma}>>^{0} &
<<C_{k\sigma} \vert C^{\dagger}_{k\sigma} >>^{0} \cr} = \frac{1}{det \hat \Omega}
\pmatrix{ 1 & - I N^{-1/2}\chi^{b}_{k \sigma}(\omega) \cr
- I N^{-1/2} \chi^{b}_{k \sigma} (\omega) & (\omega  - \varepsilon(k\sigma))  \chi^{b}_{k \sigma} (\omega)\cr}  \\
\end{equation}
where
$$ det \hat \Omega = \omega  - \varepsilon(k\sigma) - I^{2} N^{-1} \chi^{b}_{k \sigma} (\omega)$$
Let us write down explicitly the diagonal matrix elements
$G^{11}_{0}$ and $G^{22}_{0}$
\begin{equation} \label{eq90}
 <<a_{k \sigma}  \vert a^{\dagger}_{k \sigma}>>^{0} =
 (det \hat \Omega)^{-1} = (\omega  - \varepsilon(k\sigma) - I^{2} N^{-1} \chi^{b}_{k \sigma} (\omega))^{-1}
\end{equation}
The corresponding GF for the scattering regime are given by Eq.(
\ref{eq52}). As it follows from  Eqs.( \ref{eq52}) and  (
\ref{eq90}) the mean-field GF $<<a_{k \sigma}  \vert
a^{\dagger}_{k \sigma}>>^{0}$ in the bound-state regime has a
very nontrivial structure which is quite different from the
scattering-state regime form. This was achieved by a suitable
reconstruction of the generalized mean field and   by a
sophisticated redefinition of the relevant irreducible Green
functions!
We have also
\begin{equation} \label{eq91}
<<C_{k \sigma}  \vert C^{\dagger}_{k \sigma}>>^{0} =
(det \hat \Omega)^{-1} (\omega  - \varepsilon(k\sigma))(\omega  - \varepsilon(k\sigma) - det \hat \Omega))\frac{N}{I^{2}}
\end{equation}
It   follows from   Eq.( \ref{eq90}) that the quasiparticle
spectrum of the electron-magnon bound states in the generalized
mean field renormalization are determined by the equation
\begin{equation} \label{eq91a}
E_{k \sigma} = \varepsilon(k\sigma) + I^{2} N^{-1} \chi^{b}_{k \sigma} (E_{k \sigma} )
\end{equation}
The bound polaron-like electron-magnon energy spectrum consists
of two branches for any electron spin projection. At the
so-called "atomic limit" ( when $\epsilon_{k} = 0$ ) and in the
limit $k \rightarrow 0$, $\omega \rightarrow 0$ we obtain the
exact analytical representation for the single-particle GF of the
form
\begin{equation} \label{eq92}
<<a_{k \sigma}  \vert a^{\dagger}_{k \sigma}>>^{0} =
\frac{S + z_{\sigma} S_{z}}{2S + 1} ( \omega + IS)^{-1} +
\frac{S - z_{\sigma} S_{z}}{2S + 1} ( \omega - I(S + 1))^{-1}
\end{equation}
Here the notation  $S$   and $S_{z} =
\frac{<S^{z}_{0}>}{\sqrt{N}}$ means the spin-value and
magnetization, respectively.
This result was derived previously in paper.~\cite{auslend83} \\
However, our approach is the   closest to the seminal paper of
Shastry and Mattis~\cite{mattis81}, where the  Green function
treatment  of the magnetic polaron problem was formulated for
zero temperature. Our generalized mean-field solution is reduced
exactly to the Shastry-Mattis result if we put in our expression
for the spectrum, Eq.( \ref{eq91}), the temperature $ T = 0$
\begin{equation} \label{eq93}
<<a_{k \sigma}  \vert a^{\dagger}_{k \sigma}>>^{0}\vert_{T=0} =
\{ \omega - \varepsilon ( k\sigma) -
\delta_{\sigma\downarrow} 2I^{2}S \frac{\Lambda_{k\sigma}(\omega)}{(1 - I \Lambda_{k\sigma}(\omega))}\}^{-1}
\end{equation}
We can see that the magnetic polaron states are formed for
antiferromagnetic $s-d$ coupling ( $ I < 0$) only when there is a
lowering of the band of the uncoupled itinerant charge carriers
due to the effective attraction of the carrier and magnon.\\ The
derivation of Eq.( \ref{eq93}) was carried out for   arbitrary
interrelations between the s-d model parameters. Let us consider
now the two limiting cases where analytical calculations are
possible.
\begin{itemize}
\item[(i)]
a wide-band semiconductor ( $|I|S \ll W $) \\
\begin{eqnarray} \label{eq94}
E_{k  \downarrow}  \simeq \epsilon _{k} +
I \frac{S (S + S_{z} +1)   + S_{z} (S - S_{z} +1 )}{2S }   + \\
\frac{(-I)}{N}\sum_{q}   \frac{( \epsilon_{k-q} - \epsilon _{k} +
2I (S - S_{z} ))}{(\epsilon_{k-q} - \epsilon _{k} + 2IS_{z}  )}   \frac{ <S^{+}_{q}S^{-}_{-q}>}{2S} \nonumber
\end{eqnarray}
\item[(ii)]
a narrow-band semiconductor ( $|I|S \gg W $ ) \\
\begin{eqnarray} \label{eq95}
E_{k  \downarrow}  \simeq I ( S + 1 ) + \frac{2 ( S + 1 )(S + S_{z})}{( 2S + 1 )(S + S_{z} +1)}
\epsilon _{k} +
\frac{1}{N}\sum_{q} \frac{(\epsilon_{k-q} - \epsilon _{k})}{( 2S + 1 )}\frac{ <S^{+}_{q}S^{-}_{-q}>}{(S + S_{z} +1) }
\end{eqnarray}
\end{itemize}
In the above formulae the  correlation function of the
longitudinal spin components $K^{zz}_{q}$ was
omitted for the sake of simplicity. Here $W$ is the bandwidth in the limit $I = 0$.\\
Let us now consider  in  more detail the low-temperature
spin-wave limit in  Eqs.( \ref{eq94}) and  ( \ref{eq95}). In that
limit it is reasonable to suppose that $S_{z} \simeq S$. In the
spin-wave approximation we also have
$$<S^{+}_{q}S^{-}_{-q}> \simeq 2S( 1 + N(\omega (q))) $$
Thus, we obtain (c.f. Refs.~\cite{rys67,auslend83})
\begin{itemize}
\item[(i)]
a wide-band semiconductor ( $|I|S \ll W $) \\
\begin{eqnarray} \label{eq96}
E_{k  \downarrow}  \simeq \epsilon _{k} +  IS + \frac{2I^{2}S}{N}
\sum_{q} \frac{1}{(\epsilon _{k} - \epsilon_{k-q} + 2IS )} +
\frac{(-I)}{N}\sum_{q}   \frac{( \epsilon_{k-q} - \epsilon _{k} )}{(\epsilon_{k-q} - \epsilon _{k} - 2IS )}
N(\omega (q))
\end{eqnarray}
\item[(ii)]
a narrow-band semiconductor ( $|I|S \gg W $ ) \\
\begin{eqnarray} \label{eq97}
E_{k  \downarrow}  \simeq I ( S + 1 ) + \frac{2S }{( 2S + 1 )}\epsilon _{k} +
\frac{1}{N}\sum_{q} \frac{2S }{( 2S + 1 )} \frac{(\epsilon_{k-q} - \epsilon _{k})}{( 2S + 1 )}
N(\omega (q))
\end{eqnarray}
\end{itemize}
We shall now estimate the binding energy of the magnetic polaron bound state.
The binding energy of the magnetic polaron is convenient to define as
\begin{equation} \label{eq98}
\varepsilon_{B} = \varepsilon_{k\downarrow} - E_{k  \downarrow}
\end{equation}
This definition is quite natural and takes into account the fact
that in the simple Hartree-Fock approximation the spin-down band
is given by the expression
$$ \varepsilon_{k\downarrow} =  \epsilon _{k} + IS$$
Then the binding energy $\varepsilon_{B}$ behaves according to
the formula
\begin{itemize}
\item[(i)]
a wide-band semiconductor ( $|I|S \ll W $) \\
\begin{eqnarray} \label{eq99}
\varepsilon_{B} = \varepsilon^{0}_{B1} -
\frac{(-I)}{N}\sum_{q}   \frac{( \epsilon_{k-q} - \epsilon _{k} )}{(\epsilon_{k-q} - \epsilon _{k} - 2IS )}
N(\omega (q))
\end{eqnarray}
\item[(ii)]
a narrow-band semiconductor ( $|I|S \gg W $ ) \\
\begin{eqnarray} \label{eq100}
\varepsilon_{B} =  \varepsilon^{0}_{B2} -
\frac{1}{N}\sum_{q} \frac{2S }{( 2S + 1 )} \frac{(\epsilon_{k-q} - \epsilon _{k})}{( 2S + 1 )}
N(\omega (q))
\end{eqnarray}
\end{itemize}
where
\begin{eqnarray} \label{eq101}
\varepsilon^{0}_{B1} = \frac{(2I^{2}S)}{N}\sum_{q} \frac{1}{(\epsilon_{k-q} - \epsilon _{k} - 2IS )}
\simeq \frac{|I|S}{W}|I| \nonumber \\
 \varepsilon^{0}_{B2} = - I + \frac{\epsilon _{k}}{( 2S + 1 )} \simeq |I|
\end{eqnarray}
The present consideration gives the generalization of the
thermodynamic  study of the magnetic polaron. Clearly, local
magnetic order lowers the state energy of the dressed itinerant
carrier, with respect to some conduction or valence band. It is
obvious that below   $T_{N}$ of the antiferromagnet  the mobility
of spin polarons will be less than of bare
carriers~\cite{vigren73}, since they have to drag their
polarization cloud along. Experimental evidence for magnetic
polarons in concentrated magnetic semiconductors came from
optical studies of EuTe, an antiferromagnet~\cite{busch70}.
Direct measurements  of the polaron-binding energy were carried
out in Ref.~\cite{harr83}.
\section{ Quasiparticle Many-Body Dynamics and Damping of Quasiparticle States}
\subsection{Green Function Picture of Quasiparticles}
An effective way of viewing quasiparticles,  quite general and
consistent, is via the Green function scheme of many-body
theory\cite{kuzem02}  which we sketch below for completeness.\\
At sufficiently low temperatures, a few quasiparticles are
excited  and, therefore, this dilute quasiparticle gas is nearly
a noninteracting gas in the sense that the quasiparticles rarely
collide. The success of the quasiparticle concept in an
interacting many-body system is particularly striking because of
a great number of various applications. However, the range of
validity of the quasiparticle approximation, especially for
strongly interacting lattice systems, was not discussed properly
in many cases. In  systems like  simple metals,   quasiparticles
constitute long-lived, weakly interacting excitations, since
their intrinsic decay rate varies as the square of the dispersion
law, thereby justifying their use as the
building blocks for the low-lying excitation spectrum.\\
As we have mentioned earlier, to describe a quasiparticle
correctly, the Irreducible Green functions method is a very
suitable and useful tool.
\\ It is known~\cite{kuzem02}
that the GF is completely determined by the spectral weight
function $A(\omega)$. To explain this, let us remind that the GFs
are linear combinations of the time correlation functions
\begin{eqnarray}
\label{eq102} F_{AB} (t-t') = < A (t) B(t') > =  \frac {1}{2\pi}
\int^{ + \infty}_{ - \infty} d\omega
\exp [i\omega (t - t')] A_{AB} (\omega)  \\
F_{BA} (t'-t) = < B(t') A(t) > =  \frac {1}{2\pi} \int^{ +
\infty}_{ - \infty} d\omega \exp [i\omega (t'-t)] A_{BA} (\omega)
\label{eq103}
\end{eqnarray}
Here, the Fourier transforms  $A_{AB}(\omega)$  and
$A_{BA}(\omega)$ are of the form
\begin{eqnarray}
\label{eq104}
A_{BA} (\omega) = \\
Q^{-1} 2 \pi \sum_{m,n} \exp(-\beta E_{n}) (\psi^{\dagger}_{n} B
\psi_{m})
(\psi^{\dagger}_{m} A \psi_{n}) \delta ( E_{n} - E_{m} - \omega )\nonumber   \\
A_{AB} = \exp( -\beta \omega) A_{BA} ( - \omega) \label{eq105}
\end{eqnarray}
Expressions (\ref{eq104}) and (\ref{eq105}) are  spectral
representations of the corresponding time correlation functions.
The quantities $A_{AB}$ and $A_{BA}$ are  spectral
densities or spectral weight functions. \\
It is convenient to define
\begin{eqnarray}
\label{eq106} F_{BA} (0) = < B(t) A(t) > =  \frac {1}{2\pi}
\int^{ + \infty}_{ - \infty} d\omega
A (\omega)  \\
F_{AB} (0) = < A (t) B(t) > =  \frac {1}{2\pi} \int^{ + \infty}_{
- \infty} d\omega \exp (\beta \omega ) A (\omega) \label{eq.107}
\end{eqnarray}
Then, the spectral representations of the Green functions can be
expressed in the form
\begin{eqnarray}
\label{eq108}
G^{r} (\omega) = << A | B >>^{r}_{\omega} = \\
 \frac {1}{2\pi} \int^{ + \infty}_{ - \infty}
\frac {d\omega'}{ \omega - \omega' + i\epsilon}
[\exp(  \beta \omega') - \eta ] A (\omega')   \nonumber \\
G^{a} (\omega) = << A | B >>^{a}_{\omega} =  \\
\frac {1}{2\pi} \int^{ + \infty}_{ - \infty} \frac {d\omega'}{
\omega - \omega' - i\epsilon} [\exp(  \beta \omega')  - \eta ] A
(\omega')  \label{eq109} \nonumber
\end{eqnarray}
The most important practical consequence of  spectral
representations for the retarded and advanced GFs is the so-called
{\it spectral theorem}. The spectral theorem can be written as
\begin{eqnarray}
\label{eq110}
< B(t') A(t) > = \\
\nonumber
- \frac {1}{\pi} \int^{ + \infty}_{ - \infty} d\omega
\exp [i\omega (t-t')] [\exp ( \beta \omega) - \eta ]^{-1} Im G_{AB} (\omega + i\epsilon) \\
\label{eq111}
 < A(t) B(t') > = \\ \nonumber - \frac {1}{\pi} \int^{ +
\infty}_{ - \infty} d\omega \exp  (\beta \omega)  \exp [i\omega
(t-t')] [\exp ( \beta \omega) - \eta ]^{-1} Im G_{AB} (\omega +
i\epsilon)
\end{eqnarray}
Expressions (\ref{eq110}) and (\ref{eq111}) are of fundamental
importance. They directly relate the statistical averages with
the Fourier transforms of the corresponding GFs. The problem of
evaluating the latter is thus reduced to finding their Fourier
transforms  providing the practical usefulness of the Green
functions technique~\cite{tyab67,kuzem02}.\\
The spectral weight function reflects the
microscopic structure of the system under consideration.
Its Fourier transform
origination  is then the density of states that can
be reached by adding or
removing a particle of a given momentum and energy.\\
Consider a system of interacting fermions as an example. For a
noninteracting system, the  spectral weight function of the
single-particle GF
$G_{k}(\omega) = <<a_{k\sigma};a^{\dagger}_{k\sigma}>>$
has a simple peaked structure
$$A_{k}(\omega) \sim \delta (\omega - \epsilon_{k})$$~.
For an interacting system, the spectral function $ A_{k}(\omega)$
has no   such a simple peaked structure, but it   obeys the
following conditions:
$$ A_{k}(\omega) \ge 0; \quad \int A_{k}(\omega)d\omega =
<[a_{k\sigma},a^{\dagger}_{k\sigma}]_{+}> = 1$$
Thus, we can see from these expressions that  for a
noninteracting system, the sum rule is exhausted by a single
peak. A sharply peaked spectral function for an interacting
system means a long-lived single-particle-like excitation. Thus,
the spectral weight function was established here as a physically
significant attribute of   GF. The question of what is the  best
way of extracting it from a microscopic theory is the main aim of
the present theory. \\The GF for a noninteracting system is
 $G_{k}(\omega) = (\omega - \epsilon_{k})^{-1}$.
For a weakly
interacting Fermi system, we have
$G_{k}(\omega) = (\omega -
\epsilon_{k} - M_{k}(\omega))^{-1}$
where
$M_{k}(\omega)$ is the
mass operator. Thus, for a weakly interacting system, the
$\delta$-function for $ A_{k}(\omega)$ is spread into a peak of
finite width due to the   mass operator. We have
$$M_{k}(\omega \pm i \epsilon) = Re M_{k}(\omega) \mp Im
M_{k}(\omega) = \Delta_{k}(\omega) \mp \Gamma_{k}(\omega)$$
The
single-particle GF can be written in the form
\begin{equation}
\label{eq112} G_{k}(\omega) = \{\omega - [\epsilon_{k} +
\Delta_{k}(\omega)] \pm \Gamma_{k}(\omega)\}^{-1}
\end{equation}
In the weakly interacting case, we can thus find the energies of
quasiparticles by looking for the poles of single-particle GF
(\ref{eq112})
$$ \omega = \epsilon_{k} + \Delta_{k}(\omega) \pm
\Gamma_{k}(\omega)$$~.
The dispersion relation of a quasiparticle
$$ \epsilon(k) = \epsilon_{k} + \Delta_{k}[ \epsilon(k)] \pm
\Gamma_{k}[ \epsilon(k)]$$ and the lifetime $1/\Gamma_{k}$ then
reflects the inter-particle interaction. It is easy to see the
connection between the width of the spectral weight function and
decay rate. We can write
\begin{eqnarray}
\label{eq113} A_{k}(\omega) = (\exp(\beta \omega)  + 1)^{-1} (-i)
[G_{k}(\omega + i \epsilon) - G_{k}(\omega - i \epsilon)] = \\
\nonumber (\exp(\beta \omega ) + 1)^{-1} \frac
{2\Gamma_{k}(\omega)}{ [\omega - (\epsilon_{k} +
\Delta_{k}(\omega))]^{2} + \Gamma^{2}_{k}(\omega)}
\end{eqnarray}
In other words, for this case, the corresponding propagator can be
written in the form
$$ G_{k}(t) \approx \exp(-i\epsilon(k)t ) \exp( -\Gamma_{k}t)$$
This form shows under which conditions, the time-development of
an interacting system can be interpreted as the propagation of a
quasiparticle with a reasonably well-defined energy and a
sufficiently long lifetime. To demonstrate this, we consider the
following conditions:
$$\Delta_{k}[\epsilon(k)] \ll \epsilon(k); \quad
\Gamma_{k}[\epsilon(k)] \ll \epsilon(k)$$
Then we can write
\begin{equation}
\label{eq114} G_{k}(\omega) = \frac {1}{ [\omega -
 \epsilon(k)][1 - \frac {d\Delta_{k}(\omega)}{d\omega}
\vert_{\omega = \epsilon(k)}] + i\Gamma_{k}[\epsilon(k)]}
\end{equation}
where the renormalized energy of excitations is defined by
$$\epsilon(k) = \epsilon_{k} + \Delta_{k}[\epsilon(k)]$$
In this case, we have, instead of ( \ref{eq114}),
\begin{eqnarray}
\label{eq115} A_{k}(\omega) = \\ \nonumber [\exp(\beta
\epsilon(k)) + 1]^{-1} [1 - \frac {d\Delta_{k}(\omega)}{d\omega}
\vert_{\epsilon(k)}]^{-1} \frac {2\Gamma(k)}{ (\omega -
 \epsilon(k)   )^{2} + \Gamma^{2}(k)}
\end{eqnarray}
As a result, we find
\begin{eqnarray}
\label{eq116} G_{k}(t) = <<a_{k\sigma}(t);a^{\dagger}_{k\sigma}>> = \\
\nonumber = - i \theta(t) \exp(-i\epsilon(k)t) \exp( -\Gamma(k)t)
[1 - \frac {d\Delta_{k}(\omega)}{d\omega}
\vert_{\epsilon(k)}]^{-1}
\end{eqnarray}
A widely known strategy to justify this line of reasoning is the
perturbation theory. In a strongly interacted system on a lattice
with complex spectra,  the concept of a quasiparticle needs a
suitable adaptation and a careful examination. It is therefore
useful to have a workable and efficient IGF  method which, as we
have seen, permits one to determine and correctly separate the
elastic and inelastic scattering renormalizations by a correct
definition of the generalized mean field and to calculate real
quasiparticle spectra, including the damping and lifetime effects.
\subsection{ Damping of Magnetic Polaron State}
We shall now calculate the damping of the magnetic polaron state
due to the inelastic scattering effects. To obtain the Dyson
equation from  Eq.( \ref{eq84}), we have to use the relation (
\ref{eq14}). Thus , we obtain the exact Dyson equation, Eq.(
\ref{eq15}),
\begin{equation}
\label{eq117}
 \hat G(k;\omega) = \hat G^{0}(k;\omega) +   \hat G^{0}(k;\omega)  \hat
M_{k\sigma}  \hat G(k;\omega)
\end{equation}
The mass operator  has the following exact representation:
\begin{equation}
\label{eq118}
\hat M_{k\sigma} =  \pmatrix{ 0 & 0\cr
0& \Pi_{k\sigma}(\omega) \over \chi^{b}_{k \sigma} (\omega)\cr}  \\
\end{equation}
Here the notation was used
\begin{equation}
\label{eq119}
\Pi_{k\sigma}(\omega)  = \sum_{pq} \{ \frac{<<A_{p }  \vert A^{\dagger}_{q}>>^{p}}{\omega^{b}_{kp} \omega^{b}_{kq}}
+  \frac{<<A_{p }  \vert B^{\dagger}_{q}>>^{p}}{\omega^{b}_{kp} \Omega_{kq}}
+ \frac{<<B_{p }  \vert A^{\dagger}_{q}>>^{p}}{\Omega_{kp} \omega^{b}_{kq}}
+ \frac{<<B_{p }  \vert B^{\dagger}_{q}>>^{p}}{\Omega_{kp} \Omega_{kq}} \}
\end{equation}
For the single-particle GF of itinerant carriers we have
\begin{equation}
\label{eq120}
<<a_{k\sigma}\vert a^{\dagger}_{k\sigma}>> = \{ \Bigl( <<a_{k\sigma}\vert a^{\dagger}_{k\sigma}>>^{0}
\Bigr )^{-1} - \Sigma_{k \sigma}(\omega ) \}^{-1}
\end{equation}
Here the self-energy operator $\Sigma_{k \sigma}(\omega )$ was defined as
\begin{equation}
\label{eq121}
\Sigma_{k  \sigma}(\omega )  = \frac{I^{2}}{N} \frac{\Pi_{k\sigma}(\omega)}{1 -
(\chi^{b}_{k \sigma} (\omega))^{-1}\Pi_{k\sigma}(\omega)}
\end{equation}
We shall now use the exact representation, Eq.( \ref{eq121}), to
derive a suitable self-consistent approximate expression  for the
self-energy . Let us consider the GFs appearing in Eq.(
\ref{eq119}).
According to the spectral theorem, Eqs.( \ref{eq110}) and (
\ref{eq111}), it is convenient to write down the GF $<<A_{p }
\vert A^{\dagger}_{q}>>^{p}$ in the following form:
\begin{equation}
\label{eq124}
<<A_{p }  \vert A^{\dagger}_{q}>>^{p} =  \frac {1}{2\pi} \int^{ + \infty}_{ - \infty}
\frac {d\omega'}{ \omega - \omega' + i\epsilon}
[\exp(  \beta \omega') + 1 ] \int dt \exp ( i \omega't) < A^{\dagger}_{q}  A_{p }(t)>^{p}
\end{equation}
Then we obtain for the correlation function $< A^{\dagger}_{q}  A_{p }(t)>^{p}$
\begin{eqnarray}
\label{eq125}
< A^{\dagger}_{q}  A_{p }(t)>^{p}
= \frac{I^2}{N}< S^{\sigma}_{q}C^{\dagger}_{k+q-\sigma} C_{k+p-\sigma}(t)S^{-\sigma}_{-p}(t)>
+ <a^{\dagger}_{q+k-\sigma}\Phi^{\dagger}_{-q-\sigma} \Phi_{-p-\sigma}(t)a_{p+k-\sigma}(t)>
\end{eqnarray}
A further insight is gained if we select a suitable relevant
``trial" approximation for the correlation function in the r.h.s.
of (\ref{eq125}). In this paper, we show that our formulations
based on the IGF method permit one to obtain   an explicit
approximate expression for the mass operator in a self-consistent
way. It is clear  that a  relevant trial approximation for the
correlation function in (\ref{eq125}) can be chosen in various
ways. For example, a reasonable and workable one can be the
following ``mode-mode coupling approximation" which is especially
suitable for a description of two coupled subsystems.
Then the correlation function $< A^{\dagger}_{q}  A_{p }(t)>^{p}$ could now be written in the
approximate form using the following decoupling procedure (approximate trial solutions):
\begin{eqnarray}
\label{eq126}
< S^{\sigma}_{q}C^{\dagger}_{k+q-\sigma} C_{k+p-\sigma}(t)S^{-\sigma}_{-p}(t)> \simeq \delta_{q,p}
< S^{\sigma}_{q}S^{-\sigma}_{-q}(t)> < C^{\dagger}_{k+q-\sigma} C_{k+q-\sigma}(t)> \\
<a^{\dagger}_{q+k-\sigma}\Phi^{\dagger}_{-q-\sigma} \Phi_{-p-\sigma}(t)a_{p+k-\sigma}(t)> \simeq \delta_{q,p}
<\Phi^{\dagger}_{-q-\sigma} \Phi_{-q-\sigma}(t)> <a^{\dagger}_{q+k-\sigma}a_{q+k-\sigma}(t)> \nonumber
\end{eqnarray}
Here the notation was introduced
\begin{eqnarray}
\label{eq127}
<\Phi^{\dagger}_{-q-\sigma} \Phi_{-q-\sigma}(t)> =
\frac{1}{N} \sum_{pp'} J_{p}J_{p'} < \left ( ( S^{z}_{-p} )~^{ir} S^{\sigma}_{q+p} -
( S^{z}_{-(q+p)} )~^{ir} S^{\sigma}_{-p}\right )~^{ir} \nonumber \\
\left ( S^{-\sigma}_{-(p'+q)}(t) (S^{z}_{p'}(t))~^{ir} -
S^{-\sigma}_{p'}(t)( S^{z}_{-(q+p)}(t) )~^{ir} \right )~^{ir} >
\end{eqnarray}
The approximation, Eq.( \ref{eq126}), in the diagrammatic language
corresponds to   neglect  of the vertex correction, i.e., the
correlation between the propagation of the polaron and the
magnetic excitation, and the electron and magnon, respectively.
This can be performed since we already have  in our exact
expression  ( \ref{eq125}) the terms proportional to $ I^{2}$ and
$ J^{2}$. Taking into account the spectral theorem, Eqs.(
\ref{eq110}) and ( \ref{eq111}), we obtain from Eqs.( \ref{eq124})
- ( \ref{eq127})
\begin{eqnarray}
\label{eq128}
<<A_{p }  \vert A^{\dagger}_{q}>>^{p}  \simeq  \frac{I^2}{N} \delta_{q,p} \int\int
\frac {d\omega_{1} d\omega_{2}}
{\omega - \omega_{1} - \omega_{2}}
F_{1}(\omega_{1},\omega_{2})   \\
\Bigl( {-1 \over \pi} Im <<S^{-\sigma}_{-q} \vert
S^{\sigma}_{q}>>_{\omega_{1}} \Bigr)
\Bigl (  {-1 \over \pi} Im <<C_{k+q -\sigma}  \vert C^{\dagger}_{k+q -\sigma}>>_{\omega_{2}}  \Bigr )  \nonumber \\
+ \delta_{q,p}\frac{1}{N} \sum_{q'} ( J_{q'} - J_{q-q'})^{2} \int \int \int \frac {d\omega_{1} d\omega_{2} d\omega_{3}}
{\omega - \omega_{1} - \omega_{2} - \omega_{3}}
F_{2}(\omega_{1},\omega_{2},\omega_{3})  \nonumber \\
\Bigl ( {-1 \over \pi} Im <<(S^{z}_{-q'})~^{ir} \vert (S^{z}_{q'})~^{ir}>>_{\omega_{1}} \Bigr )
\Bigl ( {-1 \over \pi} Im <<S^{-\sigma}_{-(q-q')} \vert
S^{\sigma}_{q-q'}>>_{\omega_{2}} \Bigr )
\Bigl (  {-1 \over \pi} Im << a_{k+q -\sigma}  \vert a^{\dagger}_{k+q -\sigma} >>_{\omega_{3}}  \Bigr )
\nonumber
\end{eqnarray}
\begin{eqnarray}
\label{eq129}
<<B_{p }  \vert B^{\dagger}_{q}>>^{p}  \simeq  \frac{I^2}{N} \delta_{q,p} \int\int
\frac {d\omega_{1} d\omega_{2}}
{\omega - \omega_{1} - \omega_{2}}
F_{1}(\omega_{1},\omega_{2})   \\
\Bigl ( {-1 \over \pi} Im <<(S^{z}_{-q})~^{ir} \vert (S^{z}_{q})~^{ir}>>_{\omega_{1}} \Bigr )
\Bigl (  {-1 \over \pi} Im <<C_{k+q \sigma}  \vert C^{\dagger}_{k+q \sigma}>>_{\omega_{2}}  \Bigr )
\nonumber
\end{eqnarray}
where
\begin{eqnarray}
\label{eq130}
F_{1}(\omega_{1},\omega_{2}) = (1 + N(\omega_{1}) - f(\omega_{2}))  \\
F_{2}(\omega_{1},\omega_{2},\omega_{3}) = (1 + N(\omega_{1}))(1 + N(\omega_{2}) -
f(\omega_{3})) - N(\omega_{2}) f(\omega_{3}) =  \\
\label{eq131}
\Bigl ( 1 + N(\omega_{1})  \Bigr ) \Bigl ( 1 + N(\omega_{2})  \Bigr ) -
\Bigl ( 1 + N(\omega_{1}) + N(\omega_{2}) \Bigr )f(\omega_{2}) \nonumber
\nonumber
\end{eqnarray}
The functions $F_{1}(\omega_{1},\omega_{2})$,  Eq.( \ref{eq130}),
and $F_{2}(\omega_{1},\omega_{2},\omega_{3})$,   Eq.(
\ref{eq131}), represent clearly the inelastic scattering of
bosons and fermions.
For  estimation of the damping effects it is reasonably to accept
that
\begin{eqnarray}
\label{eq132}
<<A_{p }  \vert B^{\dagger}_{q}>>^{p}  \simeq   \quad  << B_{p }  \vert A^{\dagger}_{q}>>^{p}  \simeq 0
\end{eqnarray}
We have then
\begin{eqnarray}
\label{eq133}
\Pi_{k\sigma }   \simeq  \sum_{qp} \Bigl (  {<<A_{p }  \vert A^{\dagger}_{q}>>^{p} \over  \omega^{b}_{kp} \omega^{b}_{kq}}
+ {<<B_{p }  \vert B^{\dagger}_{q}>>^{p} \over  \Omega_{kp} \Omega_{kq}}   \Bigr ) \simeq \nonumber \\
\sum_{q} \Bigl (  {<<A_{q }  \vert A^{\dagger}_{q}>>^{p} \over  (\omega^{b}_{kq})^{2} }
+ {<<B_{q }  \vert B^{\dagger}_{q}>>^{p} \over  (\Omega_{kq})^{2} }   \Bigr )
\end{eqnarray}
We can see that there are two distinct contributions to the
self-energy. Putting together formulae  ( \ref{eq128}) - (
\ref{eq133}), we arrive at the following formulae for both the
contributions
\begin{eqnarray}
\label{eq134}
\Pi^{I}_{k\sigma } = \frac{I^2}{N} \sum_{q} \int\int
\frac {d\omega_{1} d\omega_{2}}
{\omega - \omega_{1} - \omega_{2}}
F_{1}(\omega_{1},\omega_{2})   \\
\{ \frac{1}{(\omega^{b}_{kq})^{2}}  \Bigl( {-1 \over \pi} Im <<S^{-\sigma}_{-q} \vert
S^{\sigma}_{q}>>_{\omega_{1}} \Bigr)
\Bigl (  {-1 \over \pi} Im <<C_{k+q -\sigma}  \vert C^{\dagger}_{k+q -\sigma}>>_{\omega_{2}}  \Bigr )  \nonumber \\
+ \frac{1}{(\Omega_{kq})^{2}} \Bigl ( {-1 \over \pi} Im <<(S^{z}_{-q})~^{ir} \vert (S^{z}_{q})~^{ir}>>_{\omega_{1}} \Bigr )
\Bigl (  {-1 \over \pi} Im <<C_{k+q \sigma}  \vert C^{\dagger}_{k+q \sigma}>>_{\omega_{2}}  \Bigr )  \}
\nonumber \\
\label{eq135}
\Pi^{J}_{k\sigma } = \frac{1}{N} \sum_{qq'} ( J_{q'} - J_{q-q'})^{2} \int \int \int \frac {d\omega_{1} d\omega_{2} d\omega_{3}}
{\omega - \omega_{1} - \omega_{2} - \omega_{3}}
F_{2}(\omega_{1},\omega_{2},\omega_{3}) \frac{1}{(\omega^{b}_{kq})^{2}}  \\
\Bigl ( {-1 \over \pi} Im <<(S^{z}_{-q'})~^{ir} \vert (S^{z}_{q'})~^{ir}>>_{\omega_{1}} \Bigr )
\Bigl ( {-1 \over \pi} Im <<S^{-\sigma}_{-(q-q')} \vert
S^{\sigma}_{q-q'}>>_{\omega_{2}} \Bigr )
\Bigl (  {-1 \over \pi} Im << a_{k+q -\sigma}  \vert a^{\dagger}_{k+q -\sigma} >>_{\omega_{3}}  \Bigr ) \nonumber
\end{eqnarray}
Equations  (\ref{eq117}), (\ref{eq118}),  (\ref{eq134}),  and
(\ref{eq135}) constitute a closed self-consistent system of
equations for the single-electron GF of the $s-d$ model  in the
bound state regime. This system of equations is much more
complicated then the corresponding system of equations for the
scattering states. We can see that to the extent that the spin and
fermion degrees of freedom can be factorized as in Eq.(
\ref{eq126}), the self-energy operator can be expressed in terms
of the initial GFs self-consistently. It is clear that this
representation does not depend on any assumption about the
explicit form of the spin and fermion GFs in the r.h.s. of
Eqs.( \ref{eq134}) and  ( \ref{eq135}). \\
Let us first consider the so-called  \emph{"static"} limit. The
thorough discussion of this approximation was carried out in
Ref.~\cite{rys67}. We just show below that a more general form of
this approximation follows directly from our formulae. The
contributions of the GFs, Eqs.( \ref{eq128}) and  ( \ref{eq129}),
are then
\begin{eqnarray}
\label{eq136}
<<A_{p }  \vert A^{\dagger}_{q}>>^{p}  \simeq  \frac{I^2}{N} \delta_{q,p} \int\int
\frac {d\omega_{1} d\omega_{2}}
{\omega - \omega_{1} - \omega_{2}}
F_{1}(\omega_{1},\omega_{2})   \\
\Bigl( {-1 \over \pi} Im <<S^{-\sigma}_{-q} \vert
S^{\sigma}_{q}>>_{\omega_{1}} \Bigr)
\Bigl (  {-1 \over \pi} Im <<C_{k+q -\sigma}  \vert C^{\dagger}_{k+q -\sigma}>>_{\omega_{2}}  \Bigr )  \nonumber \\
+ \delta_{q,p}\frac{1}{N} \sum_{q'} ( J_{q'} - J_{q-q'})^{2} <(S^{z}_{q'})~^{ir}  (S^{z}_{-q'})~^{ir}>
\int \int  \frac {d\omega_{1} d\omega_{2} }
{\omega - \omega_{1} - \omega_{2} }
F_{1}(\omega_{1},\omega_{2})  \nonumber \\
\Bigl ( {-1 \over \pi} Im <<S^{-\sigma}_{-(q-q')} \vert
S^{\sigma}_{q-q'}>>_{\omega_{1}} \Bigr )
\Bigl (  {-1 \over \pi} Im << a_{k+q -\sigma}  \vert a^{\dagger}_{k+q -\sigma} >>_{\omega_{2}}  \Bigr )
\nonumber
\end{eqnarray}
\begin{eqnarray}
\label{eq137}
<<B_{p }  \vert B^{\dagger}_{q}>>^{p}  \simeq  \frac{I^2}{N} \delta_{q,p} <(S^{z}_{q})~^{ir}  (S^{z}_{-q})~^{ir}> \int
\frac {d\omega_{1} }
{\omega - \omega_{1}}
F_{1}(\omega_{1})   \\
\Bigl (  {-1 \over \pi} Im <<C_{k+q \sigma}  \vert C^{\dagger}_{k+q \sigma}>>_{\omega_{1}}  \Bigr )
\nonumber
\end{eqnarray}
$$F_{1}(\omega_{1}) = (1  - f(\omega_{1})) $$
In the limit of low carrier concentration it is possible to drop
the Fermi distribution function in   Eqs. ( \ref{eq128})-   (
\ref{eq135}).
In principle, we can use, in the   r.h.s.  of Eqs.
(\ref{eq130})   and (\ref{eq131}), any workable first
iteration-step form of the GF and find a solution by iteration (
see Ref.~\cite{kuzem02} ). It is most convenient to choose, as
the first iteration step, the following simple one-pole
expressions:
\begin{eqnarray}
\label{eq138}
{-1 \over \pi}Im <<S^{-\sigma}_{-p}a_{k+q+p -\sigma} \vert S^{\sigma}_{p}a^{\dagger}_{k+q+p -\sigma}>>_{\omega}
= <S^{-\sigma}_{-p} S^{\sigma}_{p}> \delta (\omega + z_{\sigma}\omega_{p} -\varepsilon(k+q+p -\sigma)),  \nonumber \\
{-1 \over \pi} Im <<(S^{z}_{-p})~^{ir} a_{k+q+p \sigma} \vert (S^{z}_{p})~^{ir}a^{\dagger}_{k+q+p \sigma}>>_{\omega} =
= <(S^{z}_{-p})~^{ir}  (S^{z}_{p})~^{ir}> \delta (\omega  -\varepsilon(k+q+p -\sigma)), \nonumber \\
{-1 \over \pi}Im <<S^{-\sigma}_{-q} \vert S^{\sigma}_{q}>>_{\omega}
= - z_{\sigma} 2 <S_{z} > \delta (\omega + z_{\sigma}\omega_{q} ), \nonumber \\
{-1 \over \pi}Im <<a_{k+q+p -\sigma} \vert a^{\dagger}_{k+q+p -\sigma}>>_{\omega} =
\delta (\omega  -\varepsilon(k+q -\sigma))
\end{eqnarray}
Using Eqs.(\ref{eq124}) - (\ref{eq130}) in (\ref{eq118}) we
obtain the self-consistent approximate expression for the
self-energy operator  ( the self-consistency means that we
express approximately the self-energy operator in terms of the
initial GF,  and, in principle, one can obtain the required
solution by a suitable iteration procedure )
\begin{eqnarray}
\label{eq139}
\Sigma_{k\sigma}(\omega )  \simeq \frac{2 I^2 <S^{z}_{0}>}{N^{3/2}}\sum_{qp}
\frac{\delta_{\sigma \downarrow} + N(\omega_{q})}{(\omega^{b}_{k,q})^{2}} \Bigl (
 \frac{< S^{\sigma}_{p}S^{-\sigma}_{-p} >}{\omega + z_{\sigma}(\omega_{q} -  \omega_{p}) -\varepsilon(k+q-p -\sigma)} +
\frac{<(S^{z}_{-p})~^{ir}  (S^{z}_{p})~^{ir}>}{\omega + z_{\sigma}\omega_{q} -\varepsilon(k+q+p -\sigma)} \Bigr ) \nonumber \\
+ \frac{ I^2 }{N} \sum_{qp} \int d\omega'\frac{(1 + N(\omega'))}{(\Omega_{k,q})^{2}}
\Bigl ( {-1 \over \pi} Im <<(S^{z}_{q})~^{ir} \vert (S^{z}_{-q})~^{ir}>>_{\omega'}  \Bigr ) \nonumber \\
\Bigl ( \frac{< S^{-\sigma}_{-p} S^{\sigma}_{p}>}{\omega - \omega' + z_{\sigma} \omega_{q}   -\varepsilon(k+q+p -\sigma)} +
\frac{<(S^{z}_{-p})~^{ir}  (S^{z}_{p})~^{ir}>}{\omega - \omega' -\varepsilon(k+q+p \sigma)} \Bigr )
\end{eqnarray}
Here we write down for brevity the contribution of the $s-d$
interaction to the inelastic scattering only. For the spin-wave
approximation and low temperatures we get
\begin{equation}
\label{eq140}
\Sigma_{k \downarrow}(\omega )  \simeq \frac{(2 S I)^2 }{N}\sum_{qp}
\frac{1}{(\omega^{b}_{k,q})^{2}}
\frac{N(\omega_{p})(1 + N(\omega_{q}))}
{\omega - (\omega_{q} -  \omega_{p}) -\varepsilon(k+q-p  \downarrow)}
\end{equation}
Using the self-energy $\Sigma_{k\sigma}(\omega ) $ it is possible
to calculate the energy shift $\Delta_{k\sigma}(\omega ) = Re
\Sigma_{k\sigma}(\omega ) $ and damping $\Gamma_{k\sigma}(\omega
) =     - Im \Sigma_{k\sigma}(\omega )$ of the itinerant carrier
 in the bound state regime.
As it follows from Eq.(\ref{eq140}), the damping of the magnetic
polaron state arises from the combined processes of absorption
and emission of magnons with different energies $ (\omega_{q} -
\omega_{p})$.
Then the  real and imaginary  part  of self-energy give the
effective mass, lifetime and mobility    of the itinerant charge
carriers
\begin{eqnarray}
\label{eq141}
\frac{m^{*}}{m} =
1 - [ Re \frac{\partial  \Sigma_{k\sigma}}{\partial  \varepsilon(k\sigma)}]|_{\varepsilon(k\sigma)= \epsilon_{F}}\\
\varrho = (\frac{m}{ne^{2}})(\frac{1}{\tau}); \quad \frac{1}{\tau} =
Im \Sigma_{k\sigma} (\varepsilon(k\sigma))|_{\varepsilon(k\sigma)= \epsilon_{F}}
\end{eqnarray}
%
%%%%%%%%%%%%%%%%%%%%%%%%%%%%%%%%%%%%%%%%%%%%%%%%%%%%%%%%%%%%%%%%%%%%%%%%%%%%%%%
%
\section{Conclusions}
In summary, we have presented an analytical approach to treating
the charge quasiparticle dynamics of the  spin-fermion ($s-d$)
model  which provides a basis for description of the physical
properties of magnetic and diluted magnetic semiconductors. We
have investigated the mutual influence of the  $s-d$ and direct
exchange effects on   interacting systems of itinerant carriers
and localized spins.  We set out the theory as follows. The
workable and self-consistent IGF approach to the decoupling
problem for the equation-of-motion method for double-time
temperature Green functions has been used. The main achievement
of this formulation is the derivation of the Dyson equation for
double-time retarded Green functions instead of causal ones. That
formulation permits one to unify   convenient analytical
properties of retarded and advanced GF and   the formal solution
of the Dyson equation  which, in spite of the required
approximations for the self-energy, provides the correct
functional structure of single-particle GF. The main advantage of
the mathematical formalism is brought out by showing how elastic
scattering corrections (generalized mean fields) and inelastic
scattering effects (damping and finite lifetimes) could be
self-consistently incorporated in a general and compact manner.
This approach gives a workable scheme for definition of relevant
generalized mean fields written in terms of appropriate
correlators. A comparative study of real many-body dynamics of
the  spin-fermion model   is important  to characterize the true
quasiparticle excitations and the role of magnetic correlations.
It was shown that the charge and magnetic dynamics of the
spin-fermion model can be understood in terms of the combined
dynamics of itinerant carriers, and of localized spins and
magnetic correlations of various nature. The two other principal
distinctive features of our calculation were, first, the use of
correct analytic definition of the relevant generalized mean
fields  and, second, the explicit self-consistent calculation of
the charge and spin-wave quasiparticle spectra and their damping
for the two interacting subsystems. This analysis includes  the
scattering and bound state regimes that determine the  essential
physics. We demonstrated analytically, by contrasting the
scattering and bound state regime that the damping of magnetic
polaron is affected by both the s-d and direct  exchange. Thus,
the present consideration is the most complete analysis of the
scattering and bound state quasiparticle spectra of the
spin-fermion model. As it is seen, this treatment has advantages
in comparison with the standard methods of decoupling of higher
order GFs within the
equation-of-motion approach, namely, the following:\\
  At the mean-field level, the GF  one obtains, is richer
than that following from the standard procedures. The generalized
mean fields represent all elastic scattering renormalizations in
a compact form.\\
  The approximations ( the decoupling ) are introduced at
a later stage with respect to other methods,   i.e.,  only into
the rigorously obtained self-energy.\\
  The physical picture of   elastic and inelastic
scattering processes in the interacting many-particle systems is
clearly seen at every stage of calculations, which is not the
case with the standard methods of decoupling.\\
 Many   results of the previous works
are reproduced mathematically  more simply.\\
  The main advantage of the whole method is the
possibility of a {\it self-consistent} description of
quasiparticle spectra and their damping in a unified and coherent
fashion.
 Thus, this  picture of an interacting spin-fermion system  on a
lattice is far richer and gives more possibilities for analysis
of phenomena which can actually take place. In this sense, the
approach we suggest produces a more advanced physical picture of
the quasiparticle many-body dynamics. We have attempted to keep
the mathematical complexity within reasonable bounds by
restricting the discussion, whenever possible, to the minimal
necessary formalization. Our main results reveal the fundamental
importance of the adequate definition of generalized mean fields
at finite temperatures  which results in a deeper insight into
the nature of the bound and scattering quasiparticle states of the
correlated lattice fermions and spins. The key to understanding
of the formation of magnetic polaron in magnetic semiconductors
lies in the right description of the generalized mean fields for
coupled spin and charge subsystems. Consequently, it is crucial
that the correct functional structure of generalized mean fields
is calculated in a closed and compact form. The essential new
feature of our treatment is that it takes account  the fact that
the charge carrier  operators $( a^{\dagger}_{k\sigma},
a_{k\sigma} )$ should be treated on the equal footing with the
complex "spin-fermion" operators $(C^{\dagger}_{k\sigma},
C_{k\sigma} )$. The solution thus obtained agrees with that
obtained in the seminal paper of Shastry and
Mattis~\cite{mattis81},
where  an approach limited to zero temperature was used.\\
Finally, we wish to emphasize a broader relevance of the results
presented here to other complex magnetic materials. The detailed
consideration of the state of itinerant charge carriers in DMS
along this line will be considered separately.
%\bibliographystyle{apsrev}
%\bibliographystyle{revtex}
\bibliography{mpkuz}
%
%\begin{thebibliography}{99}
%%%%%%%%%%%%%%%%%%%%%%%%%%%
%
%
%\bibitem[\dag]{KuzEmail} Electronic address:
%kuzemsky@thsun1.jinr.ru; URL: http://thsun1.jinr.ru/~kuzemsky
%
%
%
%%%%%%%%%%%%%%%%%%%%%
%\end{thebibliography}
%%%%%%%%%%%%%%%%%%%%%
%
%
%
%%%%%%%%%%%%%%
\end{document}